\documentclass[10pt]{article}
\usepackage{amsmath}

\usepackage{amssymb,color}
\newcommand{\N}{N\raise.7ex\hbox{\underline{$\circ $}}$\;$}
\renewcommand{\theequation}{\thesection.\theequation}
\numberwithin{equation}{section}

\begin{document}

\title{Peculiarities of squaring method applied to construct  solutions    \\of the Dirac, Majorana,
and Weyl equations
 }

\maketitle

\author{O.V. Veko (Kalinkovichi Gymnazium, Belarus, vekoolga@mail.ru)}

\author{ V.M. Red'kov (Institute  of Physics, NAS of Belarus, redkov@dragon.bas-net.by)}

\begin{abstract}

It is  shown that the known  method  to solve the Dirac equation by means of the squaring method,
when relying  on the  scalar function of the  form
$
\Phi = e^{-i\epsilon t} e^{ik_{1}x} e^{ik_{2}y}
 \sin (kz + \alpha)$
 leads  to a 4-dimensional space
  of  the Dirac solutions. It is shown that
so constructed basis  is equivalent to the space of the Dirac states relied on the use of quantum number
$k_{1}, k_{2}, \pm k$ and helicity operator;
linear transformations  relating these two spaces are found.
 Application of the squaring method  substantially depends on the choice of representation
for  the Dirac matrices, some features of this are considered.
 Peculiarities of applying the squaring method in Majorana representation
are investigated as well.
The constructed bases are relevant to describe
 the Casimir effect for Dirac and Weyl fields in the
domain restricted by two planes.

\end{abstract}

{\bf Pacs:} 02.30.Gp, 02.40.Ky, 03.65Ge, 04.62.+v


\section{ Introduction}

In connection with the Casimir effect  \cite{1} for the  Dirac field in the domain restricted by two planes,
 of special interest are solutions of
the Dirac equation, which have vanishing  the third projection of the
conserved current $J^{z}$ on the plane boundaries. Such solutions
are reachable when considering 4-dimensional space
\begin{eqnarray}
 \{\Psi \} = \{
\Psi _{k_{1},k_{2},k_{3}=k, \sigma } \otimes \Psi
_{k_{1},k_{2},k_{3}= - k, \sigma} \} , \label{1.1}
\end{eqnarray}
 that is the  basis of four
solutions with opposite signs  of the third projection of momentum
 $+k_{3}$ and   $-k_{3};$ $\sigma$ is referred to polarization of the states.

It is  shown that the known  method  to solve the Dirac equation trough squaring method  \cite{2}
(elaboration of such a method
to the case of electromagnetic field see in \cite{4, 5}),
when relying  on the  scalar function of the following form
\begin{eqnarray}
\Phi = e^{-i\epsilon t} e^{ik_{1}x} e^{ik_{2}y}
 \sin (kz + \alpha) \quad  \Longrightarrow \quad
\{ \Psi \} =  \{ \Psi_{1}, \Psi_{2}, \Psi_{3}, \Psi_{4}\} \qquad
 \label{1.2}
 \end{eqnarray}

\noindent
 leads  to a 4-dimensional space  of  the Dirac solutions. It is shown that
so constructed basis (\ref{1.2}) is equivalent to the space of the Dirac states (\ref{1.1});
  linear transformations  relating these two spaces  are  found.
Different  values of parameter $\alpha$ in (\ref{1.2})  determine  only different  bases in   the
same linear  space  (\ref{1.2}).

 Application of the squaring method  substantially depends on the choice of representation
for  the Dirac matrices, some features of this circumstance  are considered.

 Peculiarities of applying the squaring method in Majorana representation
are investigated. It is shown that constructed on the base of scalar functions
 $ \cos (\epsilon t - \vec{k}{\vec x} )$ and  $-i\sin
(\epsilon t - \vec{k} \vec{x}) $,  two 4-dimensional sets of real  and imaginary solutions
of the Majorana  wave equation cannot be related by any linear transformation.

General conditions for vanishing  third projection of the  current $J^{z}$
 at the boundaries of the domain between two parallel planes are formulated:
 firstly, on the  basis of plane spinor waves (\ref{1.1});
  and secondly, on the squared basis  (\ref{1.2}).
In both cases, these conditions  reduce to a linear homogeneous algebraic systems which
leads to   a 4-th
 order algebraic equation, the roots
of which are
 $e^{2ik a}$,  where $a$ is a half-distance between the planes, and $k$ stands for
 the third projection of the Dirac  particle momentum.
Each solution of this  equation, with represent a complex number
of the unit length, provides us with   a certain rule for quantization of
the third projection of  $k$.
Explicit forms of these algebraic equations  are different, however their roots must be the same.

Conditions for vanishing the current  $J^{z}$ for Weyl neutrino field
on the boundaries of the domain between two planes are examined, the problem reduce to a 2-nd
order algebraic equation.

Covariantization of the conditions for vanishing $J^{z}$ is performed.

\section{ Squaring method}

\hspace{7mm}Let us start with a special solution of the  Klein--Fock--Gordon equation
with one formal change  in  $\Phi = e^{-i\epsilon
t + \vec{k} \vec{x}}  $: namely, we will change  the factor
$
e^{+ik_{3} z}
$ by the  real-valued factor
 \begin{eqnarray}
 \sin ( k z + \gamma) \equiv \sin \varphi\; ;
  \label{alpha}
  \end{eqnarray}
 for brevity we use notation $k=k_{3}$. Thus, we start with
 \begin{eqnarray}
(i\gamma^{a} \partial _{a}  - M) (i\gamma^{a} \partial _{a}  + M)
=
(-\partial_{t}^{2} + \partial_{j}\partial_{j} - M^{2}) \; ,
\nonumber
\end{eqnarray}
\begin{eqnarray}
\Phi = e^{-i\epsilon t} e^{ik_{1} x} e^{ik_{2} y} \sin \varphi\; ,
 \epsilon^{2} - k_{1}^{2} - k_{2}^{2} - k ^{2} - M^{2} = 0 \; .
\label{2.1}
\end{eqnarray}

\noindent The $4\times4$- matrix of the four columns-solutions of the Dirac equation
is constructed in accordance with the following rule
\begin{eqnarray}
\{ \Psi_{1}, \Psi_{2}, \Psi_{3}, \Psi_{4}  \}  =
(i\gamma^{a} \partial _{t} + i \gamma^{j} \partial _{j} + M) \Phi
\; . \label{2.2}
\end{eqnarray}

In spinor basis,  at the choice  (\ref{2.1}), the  matrix  (\ref{2.2})   will be
\begin{eqnarray}
\Psi_{1} =e^{-i\epsilon t} e^{ik_{1}x} e^{ik_{ 2}y} \left |
\begin{array}{c}
M\sin \varphi  \\
0 \\
(\epsilon \sin \varphi  + i k \cos \varphi )\\
-(k_{1} + i k_{2}) \sin \varphi
\end{array} \right |,\;
\Psi_{2} = e^{-i\epsilon t} e^{ik_{1}x} e^{ik_{ 2}y} \left |
\begin{array}{c}
 0 \\
  M \sin \varphi  \\
- (k_{1} - i k_{2}) \sin \varphi \\
(\epsilon \sin \varphi  -i k \cos \varphi )
\end{array} \right |,
\nonumber
\\[3mm]
\Psi_{3} =e^{-i\epsilon t} e^{ik_{1}x} e^{ik_{ 2}y} \left |
\begin{array}{c}
(\epsilon \sin \varphi  - i k \cos \varphi) \\
(k_{1} + i k_{2}) \sin \varphi  \\
M \sin \varphi\\
0
\end{array} \right |,\;
\Psi_{4} =e^{-i\epsilon t} e^{ik_{1}x} e^{ik_{ 2}y} \left |
\begin{array}{c}
(k_{1} - i k_{2}) \sin \varphi\\
(\epsilon  \sin \varphi + i k  \cos \varphi)\\
0 \\  M \sin \varphi
\end{array} \right |.
\label{2.4}
\end{eqnarray}

To decide on the linear dependence of $\Psi_{j}$ (or not), one should
examine the following relation
$$
a_{1} \Psi_{1} + a_{2} \Psi_{2} +  a_{3} \Psi_{3} + a_{4}
\Psi_{4} =0\; ,
$$
so we get the linear homogeneous  system  with respect to
 $a_{1}, a_{2},a_{3}, a_{4}$:
\begin{eqnarray}
a_{1} M \sin \varphi +a_{3} (\epsilon \sin \varphi  - i k \cos
\varphi)+ a_{4} (k_{1} - i k_{2}) \sin \varphi=0 \; ,
\nonumber
\\
a_{2} M \sin \varphi + a_{3} (k_{1} + i k_{2}) \sin \varphi+ a_{4}
(\epsilon \sin \varphi  + i k \cos \varphi) =0 \; ,
\nonumber
\end{eqnarray}
\begin{eqnarray}
a_{1} (\epsilon \sin \varphi  + i k \cos \varphi )- a_{2} (k_{1} -
i k_{2})\sin \varphi + a_{3} M \sin \varphi =0 \; ,
\nonumber
\\
- a_{1} (k_{1} + i k_{2}) \sin \varphi +a_{2} (\epsilon \sin
\varphi  -i k \cos \varphi ) +a_{4}  M \sin \varphi = 0 \; .
\label{2.6}
\end{eqnarray}

\noindent Its determinant turns to be
$
\det \{ \Psi \} = k^{4}$.
 This means that four solutions (\ref{2.4}) of the Dirac equation
(at any $\alpha$ in the $\varphi =
kz + \alpha $) are linearly independent: they determine a 4-dimensional space.

Let us specify two possibilities for $\gamma$ in (\ref{alpha}):
\begin{eqnarray}
\gamma = 0\;, \qquad \varphi = kz, \qquad \sin \varphi = \sin kz, \qquad \cos \varphi =
\cos  kz  \; ,\hspace{30mm}
\nonumber
\\
\Psi_{1} =e^{-i\epsilon t} e^{ik_{1}x} e^{ik_{ 2}y} \left |
\begin{array}{c}
M\sin kz  \\
0 \\
(\epsilon \sin kz  + i k \cos kz )\\
-(k_{1} + i k_{2}) \sin kz
\end{array} \right |,
\Psi_{2} = e^{-i\epsilon t} e^{ik_{1}x} e^{ik_{ 2}y} \left |
\begin{array}{c}
 0 \\
  M \sin kz  \\
- (k_{1} - i k_{2}) \sin kz \\
(\epsilon \sin kz  -i k \cos kz )
\end{array} \right |,
\nonumber
\\[2mm]
\Psi_{3} =e^{-i\epsilon t} e^{ik_{1}x} e^{ik_{ 2}y} \left |
\begin{array}{c}
(\epsilon \sin kz  - i k \cos kz) \\
(k_{1} + i k_{2}) \sin kz  \\
M \sin kz\\
0
\end{array} \right |,\;
\Psi_{4} =e^{-i\epsilon t} e^{ik_{1}x} e^{ik_{ 2}y} \left |
\begin{array}{c}
(k_{1} - i k_{2}) \sin kz\\
(\epsilon  \sin kz + i k  \cos kz)\\
0 \\  M \sin kz
\end{array} \right | ,
\label{2.8}
\end{eqnarray}

\noindent
and
\begin{eqnarray}
\gamma = {\pi \over 2} \;, \qquad \varphi ' = kz -{\pi \over 2},\qquad  \sin \varphi' = -\cos kz,
\qquad \cos \varphi'  = \sin kz \; ,\hspace{30mm}
\nonumber
\\
\Psi'_{1} =e^{-i\epsilon t} e^{ik_{1}x} e^{ik_{ 2}y} \left |
\begin{array}{c}
- M  \cos kz  \\
0 \\
(- \epsilon \cos kz + i k \sin kz  )\\
(k_{1} + i k_{2}) \cos kz
\end{array} \right |,\;
\Psi'_{(2)} = e^{-i\epsilon t} e^{ik_{1}x} e^{ik_{ 2}y} \left |
\begin{array}{c}
 0 \\
 - M \cos kz
  \\
 (k_{1} - i k_{2}) \cos kz \\
(-\epsilon \cos kz  -i k \sin kz  )
\end{array} \right |,
\nonumber
\\[2mm]
\Psi'_{3} =e^{-i\epsilon t} e^{ik_{1}x} e^{ik_{ 2}y} \left |
\begin{array}{c}
(-\epsilon \cos kz  - i k \sin kz  ) \\
-(k_{1} + i k_{2})\cos kz  \\
-M \cos kz\\
0
\end{array} \right |,\;
\Psi'_{4} =e^{-i\epsilon t} e^{ik_{1}x} e^{ik_{ 2}y} \left |
\begin{array}{c}
-(k_{1} - i k_{2}) \cos kz\\
(-\epsilon  \cos kz+ i k   \sin kz  )\\
0 \\  -M \cos kz
\end{array} \right |.
\label{2.9}
\end{eqnarray}

We readily find  expressions for the following combinations  (the total factor  $e^{-i\epsilon t} e^{ik_{1}x} e^{ik_{
2}y}$ is omitted):
\begin{eqnarray}
-\Psi'_{1} + i \Psi_{1} =  e^{ikz} \left | \begin{array}{c} M
\\
0\\
\epsilon - k
\\
-(k_{1}+ik_{2})
\end{array} \right |,
\qquad -\Psi'_{2} + i \Psi_{2} = e^{ikz} \left | \begin{array}{c}
0
\\
M\\
-(k_{1}-ik_{2})\\
\epsilon + k
\end{array} \right |,
\nonumber
\\
-\Psi'_{3} + i \Psi_{3} =  e^{ikz} \left | \begin{array}{c}
\epsilon + k \\
(k_{1}+ik_{2})\\
M\\
0
\end{array} \right |,\qquad
-\Psi'_{4} + i \Psi_{4} = e^{ikz} \left | \begin{array}{c}
(k_{1}-ik_{2})\\
\epsilon - k \\
0\\
M
\end{array} \right |.
\label{2.10}
\end{eqnarray}

Note that  solutions (\ref{2.10})  exactly coincide with those obtained by applying the squaring method
when one starts with a  scalar function as the ordinary plane wave:
\begin{eqnarray}
   U =
  e^{-i\epsilon t  + i k_{1} x + i k_{2} y + i kz}  \left |
\begin{array}{cccc}
M  &  0  & (\epsilon +k) & (k_{1} - i k_{2})\\
0  &  M  & (k_{1} + i k_{2}) &  (\epsilon - k) \\
(\epsilon - k) & -(k_{1} - i k_{2}) &  M  &  0 \\
-(k_{1} + i k_{2}) &   (\epsilon + k) & 0  & M
\end{array} \right |  .
\label{2.11}
\end{eqnarray}

The rank of the matrix in (\ref{2.11}) equals to 2. Therefore,
among four solutions of the Dirac equations given by (\ref{2.11}) only two of them are linearly independent.
 For definiteness, let us chose  solutions  $U_{(1)}$ and $U_{(2)}$:
\begin{eqnarray}
U_{1} = e^{-i\epsilon t  + i k_{1} x + i k_{2} y + i kz} \left | \begin{array}{c} M  \\ 0 \\ \epsilon - k \\
-(k_{1} + i k_{2})
\end{array} \right |, \;\;
U_{2} = e^{-i\epsilon t  + i k_{1} x + i k_{2} y + i kz}  \left | \begin{array}{c} 0  \\ M \\ -(k_{1} - ik_{2}) \\
\epsilon  +  k
\end{array} \right |  .
\label{2.12}
\end{eqnarray}

\noindent It is readily checked that remaining solutions   $U_{3},U_{4}$ are expressed through  $U_{1},U_{2}$
 as follows
\begin{eqnarray}
U_{3} = {1 \over M}  \left [ \; (\epsilon + k)   \; U_{1} +
(k_{1}+ik_{2})   \; U_{2}  \right ] \; ,\quad
 U_{4} ={1 \over M} \left [  \; (k_{1}-ik_{2}) \; U_{1} +
(\epsilon - k)   \; U_{2} \right ] \; . \label{2.13}
\end{eqnarray}

Also, instead of  (\ref{2.10}),  one can  construct other  combinations
\begin{eqnarray}
-\Psi'_{1} - i \Psi_{1} = e^{-ikz} \left | \begin{array}{c} M
\\
0\\
\epsilon + k
\\
-(k_{1}+ik_{2})
\end{array} \right |,\;
-\Psi'_{2} - i \Psi_{2} = e^{-ikz} \left | \begin{array}{c} 0
\\
M\\
-(k_{1}-ik_{2})\\
\epsilon - k
\end{array} \right |,
\nonumber
\\
-\Psi'_{3} - i \Psi_{3} =  e^{-ikz} \left | \begin{array}{c}
\epsilon - k \\
(k_{1}+ik_{2})\\
M\\
0
\end{array} \right |,\;
-\Psi'_{4} - i \Psi_{4} = e^{-ikz} \left | \begin{array}{c}
(k_{1}-ik_{2})\\
\epsilon + k \\
0\\
M
\end{array} \right | .
\label{2.14}
\end{eqnarray}

\noindent These  coincide with
those obtained  by  applying the squaring method
to a  scalar function as the ordinary plane wave
with the only change $k$ into $-k$:
\begin{eqnarray}
   U '=
  e^{-i\epsilon t  + i k_{1} x + i k_{2} y - i kz}  \left |
\begin{array}{cccc}
M  &  0  & (\epsilon -k) & (k_{1} - i k_{2})\\
0  &  M  & (k_{1} + i k_{2}) &  (\epsilon + k) \\
(\epsilon + k) & -(k_{1} - i k_{2}) &  M  &  0 \\
-(k_{1} + i k_{2}) &   (\epsilon - k) & 0  & M
\end{array} \right |  .
\label{2.15}
\end{eqnarray}

\noindent Again, the rank of the matrix in (\ref{2.15}) equals 2, so only two solutions are independent:
\begin{eqnarray}
U'_{1} = e^{ -i\epsilon t  + i k_{1} x + i k_{2} y - i kz}
\left | \begin{array}{c} M  \\ 0 \\ \epsilon + k \\
-(k_{1} + i k_{2})
\end{array} \right |, \quad
U'_{(2)} = e^{-i\epsilon t  + i k_{1} x + i k_{2} y - i kz}
 \left | \begin{array}{c} 0  \\ M \\ -(k_{1} - ik_{2}) \\
\epsilon  -  k
\end{array} \right |  .
\label{2.16}
\end{eqnarray}

\noindent Solutions  $U_{(3)}$ and $U_{(4)}$
are constructed through  $U_{(1)},U_{(2)}$
 in accordance with the rules
\begin{eqnarray}
U'_{(3)} = {1 \over M}  \left [ (\epsilon - k)   \; U'_{(1)} +
(k_{1}+ik_{2})   \; U_{(2)}  \right ] \; ,\qquad
 U'_{(4)} ={1 \over M} \left [  (k_{1}-ik_{2}) \; U'_{(1)} +
(\epsilon + k)   \; U'_{(2)} \right ] \; . \label{2.17}
\end{eqnarray}

\section*{3. Solutions in the basis of  momentum 4-vector and helicity }

Solutions of the Dirac equation in Cartesian coordinates
 can be searched in the form
\begin{eqnarray}
\Psi _{\epsilon, k_{1}, k_{2}}  = e^{-i \epsilon t} \; e^{i
k_{1}x} \; e^{ik_{2}y} \; e^{ik_{3}z}
\left | \begin{array}{c}   f_{1}  \\ f_{2}  \\   f_{3}  \\
f_{4}
\end{array} \right | \; .
\label{3.2}
\end{eqnarray}


\noindent With the use of spinor basis for Dirac matrices,
one gets the linear system for   $f_{i}$:
\begin{eqnarray}
\epsilon  f_{3}  + k_{1}  f_{4}  - i  k_{2}
 f_{4}  + k_{3}  f_{3} - M
f_{1} = 0 \; ,
\nonumber
\\
\epsilon  f_{4}  +   k_{1}  f_{3}  + i k_{2}
 f_{3}  - k_{3}  f_{4} - M
f_{2} = 0 \; ,
\nonumber
\\
\epsilon  f_{1}  -  k_{1} f_{2}  +  i k_{2}
 f_{2}  -  k_{3}  f_{1} - M
f_{3} = 0 \; ,
\nonumber
\\
\epsilon  f_{2}  -  k_{1} f_{1}  - i k_{2}
 f_{1}  + k_{3} f_{2} -  M
f_{4} = 0 \; .
 \label{3.3}
\end{eqnarray}

\noindent
Let us diagonalize  the known helicity
operator
$
\Sigma =\sigma_{j}p_{j}$.
With the substitution (\ref{3.2}), from eigenvalue equation  $ \Sigma  \; \Psi = p \; \Psi$ we
arrive at
\begin{eqnarray}
 k_{1}  f_{2}  -   ik_{2}  f_{2}  +  k_{3}  f_{1} =  p f_{1}  ,
 \nonumber
 \\
k_{1}  f_{1} \; +\;  ik_{2}  f_{1} \; -  k_{3} f_{2} =  p f_{2} \;
,
\nonumber
\\
k_{1}  f_{4}  -  ik_{2}  f_{4} +  k_{3} f_{3} =  p f_{3}
 ,\nonumber
 \\
k_{1}  f_{3}  +   ik_{2}  f_{3}  -  k_{3} f_{4} =  p f_{4}
.
 \label{3.5}
 \end{eqnarray}

\noindent Considering eqs.    (\ref{3.3})  and     (\ref{3.5}) jointly,
we obtain  the system (note that  $p^{2}  = \epsilon ^{2}  - M^{2}$)
\begin{eqnarray}
\epsilon  f_{3}  + p f_{3}   - M  f_{1} = 0 \; ,\;\; \epsilon
f_{4}  + p  f_{4}  - M \; f_{2} = 0 \; ,
\nonumber
\\
\epsilon  f_{1}  -  p  f_{1}   - M  f_{3} = 0 \; , \;\; \epsilon
f_{2}   - p  f_{2}  -  M \; f_{4} = 0 \; .
\nonumber
\label{3.6}
\end{eqnarray}

\noindent This system  results in
two values for  $p$ and corresponding restrictions on
$f_{i}$:
\begin{eqnarray}
 f_{3} =
{\epsilon - p \over M } f_{1} \;  , \quad   f_{4} = {\epsilon- p
\over M }  f_{2}\;  . \label{3.7}
\end{eqnarray}

\noindent
Allowing for  (\ref{3.7}), from  (\ref{3.5}) one gets  the system for
 $f_{1}$ and $ f_{2}$:
\begin{eqnarray}
 (  k_{3} - p  )  f_{1} +(\;   k_{1}   -i
k_{2} ) \; f_{2} \;   = 0 \; ,
\nonumber
\\
 ( k_{3}  +p )  f_{2}  -   (\;  k_{1}   +i
k_{2}  \;) f_{1} \;   = 0 \; .
 \label{3.8}
 \end{eqnarray}

\noindent
 Further we obtain
\begin{eqnarray}
f_{2} = - {i \over ik_{1} + k_{2}} ( k_{3}  -  p ) f_{1}= {k_{1} +
i k_{2} \over k_{3} + p} f_{1}  .
\nonumber
\end{eqnarray}

\noindent
Thus, two independent solutions  (we will mark them by $\alpha$ and $\beta$) at the fixed $\vec{k}$ are
 (further let    $f_{1} = 1$ and $p= + \sqrt{\epsilon^{2} - M^{2}}$)
\begin{eqnarray}
(\alpha),  \quad   {\epsilon - p \over M } = \alpha \;,
\quad f_{1} = 1, \quad f_{2 } = {k_{1} + i k_{2} \over
k_{3} + p}  =s ,
\nonumber
\\
(\beta),  \quad  {\epsilon+  p \over M }   = \beta  \;,
\qquad f_{1} = 1,  \quad f_{2 } = {k_{1} + i k_{2} \over
k_{3} - p}  = t  ;
 \label{3.9}
\end{eqnarray}

\noindent these may be presented in a shorter form
\begin{eqnarray}
\Psi _{(\alpha)}  = e^{-i \epsilon t}  e^{i k_{1}x}  e^{ik_{2}y}
e^{ik_{3}z}
\left | \begin{array}{r}   1 \\ s  \\   \alpha   \\
\alpha s
\end{array} \right | \; ,\quad
\Psi _{(\beta)}  = e^{-i \epsilon t} e^{i k_{1}x} e^{ik_{2}y}
e^{ik_{3}z}
\left | \begin{array}{c}   1  \\ t  \\   \beta   \\
\beta t
\end{array} \right | \; ,
\label{3.10}
\end{eqnarray}

\section{ Relationships  between two bases: those obtained from squaring method and
the momentum-helicity solutions}

Four types of the different  solutions of the Dirac equation
can be constructed be the method of separation of variables
(we take  $k_{3}=k$ and $k_{3}=-k$)
\begin{eqnarray}
\Phi_{1} = \Psi _{(\alpha)}(k)  = e^{ikz}
\left | \begin{array}{c}  1  \\ {k_{1} + i k_{2} \over k + p}   \\   \alpha   \\
\alpha {k_{1} + i k_{2} \over k + p}
\end{array} \right | ,
\Phi_{2} = \Psi _{(\alpha)}(-k)  = e^{-ikz}
\left | \begin{array}{c}  1  \\ {k_{1} + i k_{2} \over -k + p}  \\   \alpha   \\
\alpha {k_{1} + i k_{2} \over -k + p}
\end{array} \right | ,
\nonumber
\end{eqnarray}
\begin{eqnarray}
\Phi_{3} = \Psi _{(\beta)} (k) =  e^{ikz}
\left | \begin{array}{c}  1   \\ {k_{1} + i k_{2} \over k - p}   \\   \beta   \\
\beta {k_{1} + i k_{2} \over k - p}
\end{array} \right | , \Phi_{4}= \Psi _{(\beta)} (-k) =  e^{-ikz}
\left | \begin{array}{c}  1   \\ {k_{1} + i k_{2} \over -k - p}  \\   \beta   \\
\beta {k_{1} + i k_{2} \over -k - p}
\end{array} \right |  .
\label{4.1}
\end{eqnarray}

\noindent
and they must be related to four squaring  solutions in Section {\bf 2}.

By physical reason, we should expect existence of the following linear  expansions
\begin{eqnarray}
U_{1} = a \Phi_{1} + b \Phi_{3}, \qquad U_{2} = c \Phi_{1} +d
\Phi_{3} \; . \label{4.2a}
\end{eqnarray}

Evidently, if $a,b,c,d$ are known, one can derive
\begin{eqnarray}
U_{(3)} = {1 \over M}   [ (\epsilon + k)   \; U_{(1)} +
(k_{1}+ik_{2})   \; U_{(2)}   ]
\nonumber
\\
=
 {1 \over M}   [ (\epsilon + k)   (a \Phi_{1} + b \Phi_{3})
+ (k_{1}+ik_{2})    (c \Phi_{1} +d \Phi_{3})   ] \; ;
\label{4.2b}
\end{eqnarray}
\begin{eqnarray}
 U_{(4)} ={1 \over M}  [  (k_{1}-ik_{2}) \; U_{(1)} +
(\epsilon - k)  ; U_{(2)} ]
\nonumber
\\
= {1 \over M} [  (k_{1}-ik_{2})  (a \Phi_{1} + b \Phi_{3})
+ (\epsilon - k)   (c \Phi_{1} +d \Phi_{3})  ]\; .
\label{4.2c}
\end{eqnarray}

Analogously, there must exist expansions
\begin{eqnarray}
U'_{1} = a' \Phi_{2} + b' \Phi_{4}\; , \qquad U'_{2} = c' \Phi_{2}
+d' \Phi_{4} \; ,
\label{4.a}
\end{eqnarray}

\noindent
with the help of which one can derive
\begin{eqnarray}
U'_{(3)} ={1 \over M }   [ (\epsilon - k)   (a' \Phi_{2} + b' \Phi_{4})
+ (k_{1}+ik_{2})   \; U_{(2)}   ]\;  ,
\label{4.3b}
\\
 U'_{(4)} =
{1 \over M}  [  (k_{1}-ik_{2}) (a' \Phi_{2} + b' \Phi_{4}) +
(\epsilon + k)  (c' \Phi_{2} +d \Phi_{4})  ]\;
 .
\label{4.3c}
\end{eqnarray}

In turn, after that, one can derive the next expansions
by the  formulas
\begin{eqnarray}
\qquad  \Psi_{j} =  {1 \over 2i} (U_{j} - U'_{j} ) \; , \qquad
\Psi'_{j} = - {1 \over 2} (U_{j} + U'_{j} ) \; . \label{4.4}
\end{eqnarray}

Let us consider the first relation  in (\ref{4.2a}),
$
U_{1} = a \Phi_{1} + b \Phi_{3}\; ;
$
 explicitly it reads
\begin{eqnarray}
\left | \begin{array}{c}
M \\
0 \\
\epsilon - k\\
-(k_{1} +ik_{2})
\end{array} \right | = a
\left | \begin{array}{c}
1 \\
{k_{1} +ik_{2} \over k +p}\\
\alpha \\
\alpha {k_{1} +ik_{2} \over k+p}
\end{array}\right | + b
\left | \begin{array}{c}
1 \\
{k_{1} +ik_{2} \over k -p}\\
\beta \\
\beta {k_{1} +ik_{2} \over k-p}
\end{array}\right |;
\label{4.5a}
\end{eqnarray}

\noindent from whence it follows
\begin{eqnarray}
M = a+b, \qquad 0 = {a \over k+p} + {b \over k-p}, \qquad \epsilon
- k = a \alpha + b \beta, \qquad -1 = {a \alpha \over k+p} + {b
\beta \over k-p}\;. \label{4.5b}
\end{eqnarray}
From the first and second equations we get
\begin{eqnarray}
a ={M \over 2p} (k+p)\;, \qquad  b = - {M \over 2p} (k-p)\; ;
\label{4.5c}
\end{eqnarray}

\noindent two remaining equations turn to be identities.

Now, let us consider the second equation in (\ref{4.2a}):
$
U_{2} = c \Phi_{1} + d \Phi_{3} \; ;
$
it reads explicitly
\begin{eqnarray}
\left | \begin{array}{c}
0 \\
M \\
-(k_{1} -ik_{2}) \\
\epsilon + k
\end{array} \right | = c
\left | \begin{array}{c}
1 \\
{k_{1} +ik_{2} \over k +p}\\
\alpha \\
\alpha {k_{1} +ik_{2} \over k+p}
\end{array}\right | + d
\left | \begin{array}{c}
1 \\
{k_{1} +ik_{2} \over k -p}\\
\beta \\
\beta {k_{1} +ik_{2} \over k-p}
\end{array}\right |;
\label{4.6a}
\end{eqnarray}
or
$$
0 = c+d \; , \qquad M = c {k_{1} +ik_{2}  \over k+p} +d {k_{1}+ik_{2}
\over k-p} \; ,
$$
$$
-(k_{1} - i k_{2}) = c \alpha + d \beta \; , \qquad \epsilon + k = {c
\alpha (k_{1} + ik_{2}) \over k+p} + {d \beta (k_{1}+ik_{2}) \over
k-p} \; .
$$

From the first and second equations, it follows
\begin{eqnarray}
d= -c \; , \qquad c ={M \over 2p} (k_{1} - ik_{2})\; ;
 \label{4.6c}
\end{eqnarray}

\noindent two remaining ones are identities.
Thus, we have obtained
\begin{eqnarray}
U_{1} = {M \over 2p} \left [ (p+k) \Phi_{1} + (p-k) \Phi_{3}
\right ]\,,\qquad U_{2} = {M \over 2p } \left [ (k_{1} - ik_{2})
\Phi_{1} - (k_{1}- ik_{2})  \Phi_{3} \right ] ; \label{4.7a}
\end{eqnarray}

\noindent and further
\begin{eqnarray}
U_{3} = {M  \over  2 p }  [ {p+k \over \alpha}  \Phi_{1} + {p-k
\over \beta } \Phi_{3}  ], \qquad U_{4} ={M \over 2p }
[ { k_{1} -ik_{2} \over \alpha } \Phi_{1} - {k_{1} - ik_{2} \over
\beta } \Phi_{3}  ] ; \label{4.7b}
\end{eqnarray}

\noindent the identities  $ \epsilon -p =M
/\beta, \; \epsilon + p =M / \alpha  $ should be remembered.

Now, let us consider the  equation  $ U'_{1} = a' \Phi_{2} + b'
\Phi_{4}\; ; $ it reads explicitly as
\begin{eqnarray}
 \left |
\begin{array}{c}
M \\
0 \\
\epsilon + k\\
-(k_{1} +ik_{2})
\end{array} \right | = a'
\left | \begin{array}{c}
1 \\
{k_{1} +ik_{2} \over -k +p}\\
\alpha \\
\alpha {k_{1} +ik_{2} \over -k+p}
\end{array}\right | + b'
\left | \begin{array}{c}
1 \\
{k_{1} +ik_{2} \over -k -p}\\
\beta \\
\beta {k_{1} +ik_{2} \over -k-p}
\end{array}\right |;
\label{4.8a}
 \end{eqnarray}
 what differs formally from  (\ref{4.5a})
only in the change  $k\rightarrow -k$ a presence of primes at
variables, so the result can be written at once
$$
a' = {M \over 2p} ( - k +p)\;, \;  b' = - {M \over 2p} (-
k-p).
$$

 Consider the equation  $ U'_{2} =
c' \Phi_{2} + d' \Phi_{4}\;; $ it reads
 \begin{eqnarray} \left |
\begin{array}{c}
0 \\
M \\
-(k_{1} -ik_{2}) \\
\epsilon - k
\end{array} \right | = c'
\left | \begin{array}{c}
1 \\
{k_{1} +ik_{2} \over -k +p}\\
\alpha \\
\alpha {k_{1} +ik_{2} \over -k+p}
\end{array}\right | + d'
\left | \begin{array}{c}
1 \\
{k_{1} +ik_{2} \over -k -p}\\
\beta \\
\beta {k_{1} +ik_{2} \over -k-p}
\end{array}\right |;
\label{4.9a}
\end{eqnarray}

\noindent which differs from  (\ref{4.6a}) only in notation, so its solution looks
$
d'= -c', \; c' ={M \over 2p} (k_{1} - ik_{2})$.

Thus, we have derived decompositions
\begin{eqnarray}
U'_{1} = {M \over 2p} \left [ (p-k) \Phi_{2} + (p+k) \Phi_{4}
\right ]\,,\quad
 U'_{2} = {M \over 2p } \left [ (k_{1} - ik_{2})
\Phi_{2} - (k_{1}- ik_{2})  \Phi_{4} \right ].
\nonumber
\label{4.10a}
\\
U'_{3} = {M \over 2p} \left ( {p-k \over \alpha}  \Phi_{2} + {p+k
\over \beta} \Phi_{4} \right ),\quad
 U'_{4} ={M \over 2p}  \left
( { k_{1} -ik_{2} \over \alpha } \Phi_{2} - {k_{1} - ik_{2} \over
\beta} \Phi_{4} \right ) . \label{4.10b}
\end{eqnarray}

\noindent and repeat (\ref{4.7a}), (\ref{4.7b}):
\begin{eqnarray}
U_{1} = {M \over 2p} \left [ (p+k) \Phi_{1} + (p-k) \Phi_{3}
\right ]\,,\; U_{2} = {M \over 2p } \left [ (k_{1} - ik_{2})
\Phi_{1} - (k_{1}- ik_{2})  \Phi_{3} \right ],
\nonumber
\label{4.10c}
\\
U_{3} = {M \over 2p} \left ( {p+k \over \alpha}  \Phi_{1} + {p-k
\over \beta} \Phi_{3} \right ),\;
 U_{4} ={M \over 2p}  \left
( { k_{1} -ik_{2} \over \alpha } \Phi_{1} - {k_{1} - ik_{2} \over
\beta} \Phi_{3} \right ) . \label{4.10d}
\end{eqnarray}

Next, relying on f the formulas
\begin{eqnarray}
 \Psi_{1} =  {1 \over 2i} (U_{1} - U'_{1} ) , \quad  \Psi'_{1} = - {1 \over 2} (U_{1} + U'_{1} ) \; ,
\quad
 \Psi_{2} =  {1 \over 2i} (U_{2} - U'_{2} ) , \quad  \Psi'_{2} = - {1 \over 2} (U_{2} + U'_{1} ) \; ,
\nonumber
\\
 \Psi_{3} =  {1 \over 2i} (U_{3} - U'_{3} ) , \quad  \Psi'_{3} = - {1 \over 2} (U_{3} + U'_{3} ) \; ,
\quad
 \Psi_{4} =  {1 \over 2i} (U_{4} - U'_{4} ) , \quad  \Psi'_{4} = - {1 \over 2} (U_{4} + U'_{4} ) \; ,
\label{4.11}
\end{eqnarray}

\noindent we  derive decompositions
$$
\Psi_{l} = a_{ln} \Phi_{n} \; ,\qquad  \Psi'_{l} = a'_{ln}
\Phi_{n} \; ,
$$
 where the involved matrices  are given by
\begin{eqnarray}
i a_{ij} = {M \over  4p} \left | \begin{array}{rrrr}
(p+k) & -(p-k) & (p-k)  & -(p+k) \\
(k_{1}-ik_{2}) & -(k_{1}-ik_{2}) & -(k_{1}-ik_{2}) & (k_{1}-ik_{2}) \\
\alpha^{-1} (p+k) & -\alpha^{-1}(p-k) & \beta^{-1}(p-k)  & -\beta^{-1} (p+k)  \\
\alpha^{-1} (k_{1}-ik_{2}) & -\alpha^{-1} (k_{1}-ik_{2}) &
-\beta^{-1} (k_{1}-ik_{2}) & \beta^{-1} (k_{1}-ik_{2})
\end{array} \right | ,
\label{4.13a}
\end{eqnarray}
\begin{eqnarray}
-a'_{ij} =  {M \over 4p} \left | \begin{array}{rrrr}
 (p+k)   &  (p-k)   &  (p-k )   &  (p+k)   \\
 (k_{1}- ik_{2})    &   (k_{1}- ik_{2} )   &  - (k_{1}- ik_{2}) & -   (k_{1}- ik_{2} )\\
 \alpha^{-1} (p + k )  & \alpha^{-1} (p - k)  &  \beta^{-1} (p - k)  & \beta^{-1} (p + k) \\
 \alpha^{-1} (k_{1}-ik_{2} )  & \alpha^{-1} (k_{1}- ik_{2}) &- \beta^{-1} (k_{1}- ik_{2})
 &
 -  \beta^{-1} (k_{1}- ik_{2})
 \end{array}
 \right |.
 \label{4.13b}
 \end{eqnarray}

From those we can construct  a new matrix
 \begin{eqnarray}
S_{ij} = -a'_{ij} + i a_{ij} = {M \over 2p} \left |
\begin{array}{rrrr}
 (p+k)   &  0  &  (p-k )   &  0   \\
 (k_{1}- ik_{2})    &   0  &  - (k_{1}- ik_{2}) & 0\\
 \alpha^{-1} (p + k )  & 0  &  \beta^{-1} (p - k)  & 0 \\
 \alpha^{-1} (k_{1}-ik_{2} )  & 0 &- \beta^{-1} (k_{1}- ik_{2})
 &
 0
 \end{array}
 \right | ;
 \label{4.14a}
 \end{eqnarray}

\noindent  (it corresponds to the use in the factor $e^{ikz}$ in the
 initial scalar substitution for $\Phi$)
 The rank of this matrix equals 2, and  it is responsible for transformation
$$
U_{1}, U_{2},U_{3} , U_{4} \; \stackrel{S}{\Longleftarrow }
\;\Phi_{1}, \Phi_{3} \;.
$$

Analogously, we have another combination (it corresponds to the use in the factor $e^{-ikz}$ in the
 initial scalar substitution for $\Phi$)
\begin{eqnarray}
S'_{ij} = -a'_{ij} - i a_{ij} = {M \over 2p} \left |
\begin{array}{rrrr}
 0  &  (p-k)                     &  0   &  (p+k)   \\
 0  &   (k_{1}- ik_{2} )         &  0   & -   (k_{1}- ik_{2} )\\
 0  & \alpha^{-1} (p - k)        &  0   & \beta^{-1} (p + k) \\
 0  & \alpha^{-1}(k_{1}- ik_{2}) &  0   &  -  \beta^{-1} (k_{1}- ik_{2})
 \end{array}
 \right |;
 \label{4.15a}
 \end{eqnarray}

\noindent  the rank  of this matrix equals 2, and  it corresponds to the transformation
$$
U'_{1}, U'_{2},U'_{3} , U'_{4} \ \stackrel{S}{\Longleftarrow
} \qquad \Phi_{2}, \Phi_{4} \;.
$$

Because  the determinants  do  not vanish
\begin{eqnarray}
\det (a_{ij})  = + {M\alpha^{2} k^{2} \over ip} (k_{1}-ik_{2})^{2}
\left (\alpha^{2} + \beta^{2} -2 \right ),
\nonumber
\\
\det (a'_{ij})= -{Mk^{2} \over ip } (k_{1}-ik_{2})^{2}\left
(\alpha^{2} + \beta^{2} -2 \right ) , \label{4.17}
\end{eqnarray}

\noindent inverse matrices exist and have the form
\begin{eqnarray}
[ a^{-1}_{ij} ] = {2ip \over  kM(\alpha- \beta)}= \left |
\begin{array}{rrrr}
\alpha & {\alpha  (k-p) \over k_{1}-ik_{2}} & -1  & -{ (k-p) \over k_{1}-ik_{2}}  \\
\alpha  &  -{\alpha  (k+p) \over k_{1}-ik_{2}} & -1  &  { (k+p) \over k_{1}-ik_{2}} \\
\beta &  {\beta (k+p) \over k_{1}-ik_{2}} & -1  &  -{ (k+p) \over k_{1}-ik_{2}}   \\
\beta & -{\beta  (k-p) \over k_{1}-ik_{2}} & -1  &
 { (k-p) \over k_{1}-ik_{2}}
\end{array} \right |,
\label{4.18a}
\end{eqnarray}
\begin{eqnarray}
\;[ (a')^{-1}_{ij} ]= {2ip \over  kM(\alpha- \beta)}
 =  \left |
\begin{array}{rrrr}
-\alpha & -{\alpha  (k-p) \over k_{1}-ik_{2}} & 1  & { (k-p) \over k_{1}-ik_{2}}  \\
\alpha  &  -{\alpha  (k+p) \over k_{1}-ik_{2}} & -1  &  { (k+p) \over k_{1}-ik_{2}} \\
-\beta & - {\beta (k+p) \over k_{1}-ik_{2}} & 1  &  { (k+p) \over k_{1}-ik_{2}}   \\
\beta & -{\beta  (k-p) \over k_{1}-ik_{2}} & -1  &
 { (k-p) \over k_{1}-ik_{2}}
\end{array} \right |.\quad
\label{4.18b}
\end{eqnarray}

Evidently, we have inverse transformations relating $\Psi_{j}$
and   $\Psi'_{j}$:
$$
\Psi'_{j} = a'_{jk} \Phi_{k} =   [ a'_{jk} (a^{-1}) _{kl} ]\;
\Psi_{k}\; ,
$$
\begin{eqnarray}
 a'_{jk}   (a^{-1}) _{kl} =
 {i \over 2 k(\alpha- \beta)}\left | \begin{array}{cccc}
(p+k)\alpha & -{(p-k)^{2}\alpha \over k_{1}-ik_{2}} & -(p-k)  & {p^{2}-k^{2} \over k_{1}-ik_{2}}  \\
(k_{1}-ik_{2})\alpha &  -(k+p)\alpha & (k_{1}-ik_{2})  & -(k+p) \\
{(k+p)\beta \over \alpha} & {(p^{2}-k^{2})\beta \over (k_{1}-ik_{2})\alpha}  & -{p-k \over \beta}  &   -{(p+k)^{2}\over (k_{1}-ik_{2})\beta}\\
{(k_{1}-ik_{2})\beta \over \alpha} & {(p-k)\beta \over \alpha} &
{k_{1}-ik_{2} \over \beta}  & {p-k \over \beta}
\end{array} \right |,
\label{4.19a} \end{eqnarray}
$$
\Psi_{j} = a_{jk} \Phi_{k} =   [ a_{jk} (a'^{-1}) _{kl} ]\;
\Psi'_{k}\; ,
$$
\begin{eqnarray}
[ a_{jk}   (a'^{-1}) _{kl} ] = {1\over  k(\alpha- \beta)} \left |
\begin{array}{cccc}
-(p+k)\alpha & -{(p-k)^{2}\alpha \over k_{1}-ik_{2}} & (p-k)  & {p^{2}-k^{2} \over k_{1}-ik_{2}}  \\
(k_{1}-ik_{2})\alpha &  (k+p)\alpha & (k_{1}-ik_{2})  & (k+p) \\
{-(k+p)\beta \over \alpha} & {(p^{2}-k^{2})\beta \over (k_{1}-ik_{2})\alpha}  & {p-k \over \beta}  &   -{(p+k)^{2}\over (k_{1}-ik_{2})\beta}\\
{(k_{1}-ik_{2})\beta \over \alpha} & -{(p-k)\beta \over \alpha} &
{k_{1}-ik_{2} \over \beta}  & -{p-k \over \beta}
\end{array} \right | .
\label{4.19b}
\end{eqnarray}

 The following conclusion can be given: the choice of  an initial
phase  $\gamma$ in the function  $\sin (kz +  \gamma )$ does not
influence on the whole structure of the space  of solutions -- it
only determines a  specific basis in the same space:
\begin{eqnarray} \sin (kz +
\gamma  ) = \cos \gamma   \;  [ \sin kz ] -  \sin \gamma  \; [- \cos
kx  ] \; ,
\nonumber
\\
\sin (kz +  \gamma  )   \qquad \Longrightarrow  \qquad
\Psi^{\gamma }_{j}= \cos \gamma  \; \Psi _{j} - \sin \gamma  \;
\Psi'_{j} \; . \label{4.20}
\end{eqnarray}

\section{ Dependence of squared solutions on the choice of Dirac
matrices,
standard basis }

Let us follows two representations for Dirac matrices
$$
\gamma^{a} \; , \quad  \Gamma^{a} = S \gamma^{a} S^{-1}.
$$
 Having applied  the squaring method in both bases, we
will obtain two sets of solutions:
\begin{eqnarray}
(i \gamma^{a} \partial_{a} + M )\Phi = \{  \Psi_{1}, \Psi_{2},
\Psi_{3}, \Psi_{4} \}  \; , \qquad
(i \Gamma^{a} \partial_{a} + M )\Phi = \{  \varphi_{1}, \varphi_{2},
\varphi_{3}, \varphi_{4} \}  \; . \label{5.3}
\end{eqnarray}

\noindent These two sets, or  two matrices, are
linked to each other according to the formula
\begin{eqnarray}
S \{  \Psi_{1}, \Psi_{2}, \Psi_{3}, \Psi_{4} \}  S^{-1} = \{
\varphi_{1}, \varphi_{2}, \varphi_{3}, \varphi_{4} \} \; .
\label{5.4}
\end{eqnarray}

This relation tells  that if at a fixed initial
scalar function $\Phi$ we have a matrix of squared solutions  with
rank 2, the corresponding matrix of solutions in any other Dirac
basis   will have also the rank 2. Analogously, if we have the
matrix of solutions of the rank 4 in one Dirac basis, we will have in
any other  Dirac basis a matrix of solutions of  the same  rank 4.

For instance, let us specify squared  solutions in the commonly
used standard representation for Dirac matrices:
\begin{eqnarray}
\gamma^{0} = \left | \begin{array}{cc}
I & 0 \\
0 & -I
\end{array} \right | , \qquad
\gamma^{i} = \left | \begin{array}{cc}
0 & \sigma^{i} \\
-\sigma^{i}  & 0
\end{array} \right | .
\nonumber
\label{5.5a}
\end{eqnarray}

\noindent
Taking into account the relation
\begin{eqnarray}
(i\gamma^{a} \partial_{a} + M) = \left | \begin{array}{cccc}
i\partial_{t}+M  & 0 & i \partial_{3} & i \partial_{1} +  \partial_{2} \\
0 & i \partial_{t}+M  & i\partial_{1} - \partial_{2} & -i \partial_{3} \\
-i \partial_{3} & -i \partial_{1} - \partial_{2}  & -i \partial_{t}+M  & 0\\
-i\partial_{1} +\partial_{2} &  i \partial_{3} & 0 &
-i\partial_{t}+M
\end{array} \right |,
\nonumber
\end{eqnarray}

\noindent and choosing
$$
\Phi = e^{-i\epsilon t} e^{ik_{1}x} e^{ik_{2} y} \sin \varphi\; ,
\; \sin \varphi = kz + \gamma
\;,
$$
 we get  an explicit form for the  matrix of solutions
\begin{eqnarray}
[ W ] = \left | \begin{array}{llll}
(\epsilon  +M) \sin \varphi  & 0 & +ik \cos \varphi & (-k_{1} +ik_{2}) \sin \varphi \\
0 & (\epsilon +M)\sin \varphi  & (-k_{1} -ik_{2}) \sin \varphi & -i k \cos \varphi \\
-ik \cos \varphi & (k_{1} -ik_{2}) \sin \varphi & (-\epsilon +M) \sin \varphi & 0 \\
(k_{1} +ik_{2}) \sin \varphi & +ik \cos \varphi & 0 & (-\epsilon +
M) \sin \varphi
\end{array}  \right |.
\end{eqnarray}

\noindent Thus, we construct  four different solutions of the Dirac
equation:
\begin{eqnarray}
 W_{1} =
\left | \begin{array}{c}
(\epsilon  +M) \sin \varphi   \\
0   \\
-ik \cos \varphi  \\
(k_{1} +i k_{2}) \sin \varphi
\end{array}  \right |, \qquad
 W_{2} =
\left | \begin{array}{c}
 0 \\
(\epsilon +M) \sin \varphi  \\
 (k_{1} -ik_{2}) \sin \varphi  \\
+ik \cos \varphi
\end{array}  \right |,
\nonumber
\\
 W_{3} =
\left | \begin{array}{c}
 +ik \cos \varphi \\
 (-k_{1} -ik_{2}) \sin \varphi  \\
 (-\epsilon +M) \sin \varphi  \\
 0
\end{array}  \right |, \qquad
 W_{4} =
\left | \begin{array}{c}
(-k_{1} +ik_{2}) \sin \varphi \\
 -i k \cos \varphi \\
 0 \\
 (-\epsilon + M) \sin \varphi
\end{array}  \right | .
\label{5.5b}
\end{eqnarray}

\noindent Due to statement  after  (\ref{5.4}), they are linearly
independent.

Two choices of the phase $\gamma $  lead to two sets of solutions:
$$
 \varphi = kz, \qquad \sin \varphi = \sin kz, \qquad \cos \varphi =
\cos  kz  \; ,
$$
\begin{eqnarray}
 W_{1} =
\left | \begin{array}{c}
(\epsilon  +M) \sin kz   \\
0   \\
-ik \cos kz \\
(k_{1} +i k_{2}) \sin kz
\end{array}  \right |, \qquad
 W_{2} =
\left | \begin{array}{c}
 0 \\
(\epsilon +M) \sin kz  \\
 (k_{1} -ik_{2}) \sin kz  \\
+ik \cos kz
\end{array}  \right |,
\nonumber
\\
 W_{3} =
\left | \begin{array}{c}
 +ik \cos kz \\
 (-k_{1} -ik_{2}) \sin kz\\
 (-\epsilon +M) \sin kz  \\
 0
\end{array}  \right |, \qquad
 W_{4} =
\left | \begin{array}{c}
(-k_{1} +ik_{2}) \sin kz \\
 -i k \cos kz \\
 0 \\
 (-\epsilon + M) \sin kz
\end{array}  \right | ;
\label{5.6a}
\end{eqnarray}
$$
\varphi ' = kz -{\pi \over 2},\qquad  \sin \varphi' = -\cos kz,
\qquad \cos \varphi'  = \sin kz \; ,
$$
\begin{eqnarray}
 W'_{1} =
\left | \begin{array}{c}
-(\epsilon  +M) \cos kz   \\
0   \\
-ik \sin kz \\
-(k_{1} +i k_{2}) \cos kz
\end{array}  \right |, \qquad
 W'_{2} =
\left | \begin{array}{c}
 0 \\
-(\epsilon +M) \cos kz  \\
 -(k_{1} -ik_{2}) \cos kz  \\
+ik \sin kz
\end{array}  \right |,
\nonumber
\\
 W'_{3} =
\left | \begin{array}{c}
 +ik \sin kz \\
 (k_{1} +ik_{2}) \cos kz\\
 (\epsilon - M) \cos kz  \\
 0
\end{array}  \right |, \qquad
 W'_{4} =
\left | \begin{array}{c}
(k_{1} - ik_{2}) \cos kz \\
 -i k \sin kz  \\
 0 \\
 (\epsilon - M) \cos kz
\end{array}  \right | .
\label{5.6b}
\end{eqnarray}

From these,  one easily constructs linear combinations referring  to
application of the squaring method for two differen choices  of a  scalar function:
\begin{eqnarray}
e^{-i\epsilon t} e^{ik_{1} x} e^{ik_{2}y}\; [-W'_{j} + i W_{j} ] = \left (i
\gamma^{a}_{stand} \partial_{a} + M \right  ) \; e^{-i\epsilon t} e^{ik_{1}
x} e^{ik_{2}y}  e^{+ikz}\; ,
\nonumber
\\[2mm]
 e^{-i\epsilon t} e^{ik_{1} x} e^{ik_{2}y}\; [ -W'_{j} - i W_{j} ] = \left
  (i \gamma^{a}_{stand} \partial_{a} + M  \right ) \; e^{-i\epsilon t} e^{ik_{1} x} e^{ik_{2}y}  e^{-ikz}\; .
\label{5.7}
\end{eqnarray}

Evidently, one can perform  the  analysis like given after
(\ref{2.11})--(\ref{2.17}) with some minor technical alterations.

\section{ Squaring method and  Majorana fermion
 }

Special interest for the method has any Majorana basis for
Dirac matrices. Let us specify one of them as follows
\begin{eqnarray}
\Psi_{M} = A \Psi _{spinor}, \quad A = {1 - \gamma^{2} \over
\sqrt{2}} , \; A^{-1} =  {1 + \gamma^{2} \over \sqrt{2}},
\quad \Gamma^{a}_{M} = A \gamma^{a} A^{-1} \; ;
\label{6.1}
\end{eqnarray}
explicitly these matrices read
\begin{eqnarray}
\gamma^{0}_{M} = + \gamma^{0} \gamma^{2} =  \left |
\begin{array}{cccc}
0 & -i & 0 & 0 \\
i & 0 & 0 & 0 \\
0 & 0 & 0 &  i  \\
0 & 0 & - i & 0
\end{array} \right |,\quad
\gamma^{1}_{M} = + \gamma^{1} \gamma^{2} =
 \left | \begin{array}{cccc}
 -i & 0& 0 & 0 \\
 0  & i & 0 & 0 \\
0 & 0 & -i  &  0  \\
0 & 0 & 0 &  i
\end{array} \right |,
\nonumber
\\
\gamma^{2}_{M} =  \gamma^{2} =
 \left | \begin{array}{cccc}
0 & 0 & 0 & i \\
0 & 0 & -i & 0 \\
0 & -i & 0 &  0  \\
i & 0 &  0  & 0
\end{array} \right |,\quad
\gamma^{3}_{M} = + \gamma^{3} \gamma^{2} =
 \left | \begin{array}{cccc}
0 & i & 0 & 0 \\
i & 0 & 0 & 0 \\
0 & 0 & 0 &  i  \\
0 & 0 &  i & 0
\end{array} \right |.\qquad
\nonumber
\end{eqnarray}

First, let us construct real and pure imaginary solutions of the Dirac
equation, staring from the Dirac solutions with n momentum-helicity  quantum numbers.
To this end, it is enough to translate the Dirac plane waves
(\ref{3.10})
of the types $\alpha, \beta$   from spinor basis to Majorana one
(\ref{6.1}), and after that we are to separate real and imaginary parts of
these solutions. In this way we will construct  wave functions for Majorana particles with different
charge parities.

Thus, we start with the plane waves of the form
\begin{eqnarray}
\Psi _{(\alpha)}  = e^{-i \epsilon t} e^{i\vec{k}\vec{x}}
\left | \begin{array}{r}   1 \\ s  \\   \alpha   \\
\alpha s
\end{array} \right | \; ,\quad \alpha = {\epsilon - p \over M } \;,
\quad s = {k_{1} + i k_{2} \over k_{3} + p} \; ,
\nonumber
\\
\Psi _{(\beta)}  = e^{-i \epsilon t} e^{i\vec{k}\vec{x}}
\left | \begin{array}{c}   1  \\ t  \\   \beta   \\
\beta t
\end{array} \right | \; , \quad \beta = {\epsilon+  p \over M }   \; ,
 \quad t = {k_{1} + i k_{2} \over k_{3} - p}  \; .
\label{6.3}
\end{eqnarray}

\noindent With the help of the matrix $A$ (\ref{6.1})
 we translate them to the  Majorana representation:
\begin{eqnarray}
\Psi _{M(\alpha)}  =  {e^{-i kx} \over \sqrt{2}} \left |
\begin{array}{r}
(1+ i \alpha s)    \\
(s-i \alpha)   \\
-i
(s+i \alpha)    \\
i(1- i \alpha s)
\end{array} \right |  ,\qquad
\Psi _{M(\beta)}  =  {e^{-i kx} \over \sqrt{2}} \left |
\begin{array}{c}
(1+ i \beta t)    \\
(t-i \beta)   \\
-i
(t+i \beta)    \\
i(1- i \beta t)
\end{array} \right |   ,
\label{6.4}
\\
\Psi^{*} _{M(\alpha)}  =  {e^{+i kx } \over \sqrt{2}} \left |
\begin{array}{r}
(1- i \alpha s^{*})    \\
(s^{*}+i \alpha)   \\
i(s^{*}-i \alpha)    \\
-i(1+ i \alpha s^{*})
\end{array} \right |  ,\qquad
\Psi^{*} _{M(\beta)}  = {e^{+i kx } \over \sqrt{2}} \left |
\begin{array}{c}
(1- i \beta t^{*})    \\
(t^{*}+i \beta)   \\
i(t^{*}-i \beta)    \\
-i(1+ i \beta t^{*})
\end{array} \right |   ;
\label{6.5}
\end{eqnarray}

\noindent where  $ kx = \epsilon t - \vec{k}\;\vec{x} $. Then, from
 wave functions (\ref{6.4}), one separates real and imaginary parts:
$$
R_{M(\alpha)}={1 \over 2} (\Psi _{M(\alpha)}  + \Psi^{*}
_{M(\alpha)} )
$$
$$
=  {1 \over \sqrt{2}} \left \{  \cos kx \left | \begin{array}{r}
1+ i \alpha (s - s^{*})/2    \\
(s +s^{*} )/2   \\
 \alpha - i(s-s^{*})/2   \\
\alpha (s + s^{*} )/2
\end{array} \right | + \sin kx
\left | \begin{array}{r}
  \alpha (s+s^{*})/2    \\
-i(s-s^{*})/2 - \alpha    \\
-(s+s^{*}) /2    \\
1 -i \alpha (s - s^{*})/2
\end{array} \right |  \right \}  ,
$$
$$
I_{M(\alpha)} = {1 \over 2} (\Psi _{M(\alpha)}  - \Psi^{*}
_{M(\alpha)} )
$$
$$
=
 {1 \over \sqrt{2}} \left \{ \cos kx \left | \begin{array}{r}
 i \alpha (s + s^{*})/2    \\
(s-s^{*} )/2 -i\alpha   \\
-i(s +s^{*})/2    \\
i + \alpha (s-s^{*})/2
\end{array} \right |+ \sin kx
\left | \begin{array}{r}
- i  + \alpha (s-s^{*})/2\\
-i(s+s^{*} )/2   \\
-(s-s^{*})/2 - i\alpha   \\
-i \alpha (s+s^{*})/2
\end{array} \right | \right \},
$$

\noindent
and
$$
R_{M(\beta)} = {1 \over 2} (\Psi _{M(\beta)}  + \Psi^{*}
_{M(\beta)} )
$$
$$
=
  \cos kx \left | \begin{array}{r}
2+ i \beta (t - t^{*})    \\
t +t^{*}    \\
2 \beta - i(t-t^{*})   \\
\beta (t + t^{*} )
\end{array} \right | + \sin kx
\left | \begin{array}{r}
  \beta (t+t^{*})    \\
-i(t-t^{*}) -2 \beta    \\
-(t+t^{*})    \\
2 -i \beta (t - t^{*})
\end{array} \right |  ,
$$
$$
I_{M(\beta)} =  {1 \over 2} (\Psi _{M(\beta)}  - \Psi^{*}
_{M(\beta)} )
$$
$$
=
 \cos kx \left | \begin{array}{r}
 i \beta (t + t^{*})    \\
(t-t^{*} ) -2i\beta   \\
-i(t +t^{*})    \\
2i + \beta (t-t^{*})
\end{array} \right |+ \sin kx
\left | \begin{array}{r}
- 2i  + \beta (t-t^{*})\\
-i(t+t^{*} )   \\
-(t-t^{*}) -2 i\beta   \\
-i \beta (t+t^{*})
\end{array} \right |.
$$

Now let us detail the squaring method relied in the Majorana basis. Starting with
\begin{eqnarray}
(i\gamma_{M}^{a} \partial_{a} + M) = \left | \begin{array}{cccc}
\partial_{1} +M  & \partial_{t} - \partial_{3} & 0  & -\partial_{2} \\
-\partial_{t} -\partial_{3} & - \partial_{1} +M & \partial_{2} & 0\\
0 & \partial_{2} & \partial_{1} +M & - \partial_{t}- \partial_{3} \\
-\partial_{2} & 0  & \partial_{t} - \partial_{3} & - \partial_{1}
+M
\end{array}\right |
\nonumber
\end{eqnarray}

\noindent and taking
$
\Phi  = e^{-i\epsilon t} e^{ik_{1} x}  e^{ik_{2} y} e^{ik_{3} z}
$, we derive an explicit form for  a matrix of four solutions
\begin{eqnarray}
\{ \Psi_{j} \}  = e^{-ikx} \left | \begin{array}{cccc}
ik_{1} +M  & -i\epsilon - ik_{3} & 0  & -ik_{2} \\
i\epsilon -ik_{3} & - ik_{1} +M & ik_{2} & 0\\
0 & ik_{2} & ik_{1} +M & i\epsilon- ik_{3} \\
-ik_{2} & 0  & -i\epsilon - ik_{3} & - ik_{1} +M
\end{array}\right |.
\label{6.9b}
\end{eqnarray}

\noindent This matrix can be decomposed into real and imaginary parts, The real part reads
\begin{eqnarray}
 \{ R_{j} \} = \left | \begin {array}{cccc}
M\cos kx + k_{1} \sin kx & -\epsilon \sin kx -k_{3} \sin kx  & 0 & -k_{2} \sin kx \\
\epsilon \sin kx -k_{3} \sin kx & M\cos kx - k_{1} \sin kx & k_{2} \sin kx & 0 \\
0 & k_{2} \sin kx & M\cos kx + k_{1} \sin kx  & \epsilon \sin kx -k_{3} \sin kx\\
-k_{2} \sin kx & 0 & -\epsilon \sin kx -k_{3} \sin kx  & M\cos kx
- k_{1} \sin kx
\end{array} \right |.
\nonumber
\label{6.10a}
\end{eqnarray}

\noindent
 Because the determinant of this matrix does not vanish
 $$
\det  \{ R_{j} \}   =
(M^{2}\cos^{2}kx-(k_{1}^{2}+k_{2}^{2}-\epsilon^{2}+k_{3}^{2})\sin^{2}kx)^{2}=M^{4}
\; ;
 $$

\noindent its columns represent linearly independent solutions of
the Majorana equation.

In similar manner we consider the matrix
$\{I_{j}\}$:
\begin{eqnarray}
 [I_{j}] =
i \left | \begin{array}{cccc}
k_{1} \cos kx - M \sin kx &  -\epsilon \cos kx - k_{3} \cos kx  &  0  &-k_{2} \cos kx    \\
\epsilon \cos kx - k_{3} \cos kx &  -k_{1} \cos kx - M \sin kx & k_{2} \cos kx   &  0 \\
0& k_{2} \cos kx   &   k_{1} \cos kx - M \sin kx &  \epsilon \cos kx - k_{3} \cos kx \\
-k_{2} \cos kx &  0 & -\epsilon \cos kx - k_{3} \cos kx   & -k_{1}
\cos kx - M \sin kx
\end{array} \right |,
\nonumber
\label{6.11a}
\end{eqnarray}

\noindent its determinant is
$$
\det  \{I_{j} \}  =
((k_{1}^{2}+k_{2}^{2}-\epsilon^{2}+k_{3}^{2})\cos^{2}kx-M^{2}\sin^{2}kx)^{2}=M^{4}\; ,
 $$

\noindent therefore its columns dive linearly independent four solutions  of the Majorana equation (with
opposite charge parity).

Let us write down explicit form of four real  solutions:
\begin{eqnarray}
R_{1} = \left | \begin {array}{c}
M\cos kx + k_{1} \sin kx \\
\epsilon \sin kx -k_{3} \sin kx  \\
0 \\
-k_{2} \sin kx
\end{array} \right |, \qquad
R_{2} = \left | \begin {array}{c}
  -\epsilon \sin kx -k_{3} \sin kx   \\
 M\cos kx - k_{1} \sin kx  \\
 k_{2} \sin kx \\
0
\end{array} \right |,
\nonumber
\\
R_{3} = \left | \begin {array}{c}
 0 \\
 k_{2} \sin kx  \\
M\cos kx + k_{1} \sin kx  \\
-\epsilon \sin kx -k_{3} \sin kx
\end{array} \right |,\qquad
R_{4} = \left | \begin {array}{c}
 -k_{2} \sin kx \\
 0 \\
\epsilon \sin kx -k_{3} \sin kx\\
 M\cos kx - k_{1} \sin kx
\end{array} \right | ,
\label{6.12}
\end{eqnarray}

\noindent and four imaginary solutions
\begin{eqnarray}
I_{1}=i \left | \begin{array}{c}
k_{1} \cos kx - M \sin kx   \\
\epsilon \cos kx - k_{3} \cos kx  \\
0  \\
-k_{2} \cos kx
\end{array} \right |,\qquad
I_{2}=i \left | \begin{array}{c}
  -\epsilon \cos kx - k_{3} \cos kx     \\
  -k_{1} \cos kx - M \sin kx  \\
 k_{2} \cos kx   \\
  0
\end{array} \right |,
\nonumber\\
I_{3}=i \left | \begin{array}{c}
  0    \\
k_{2} \cos kx    \\
   k_{1} \cos kx - M \sin kx \\
 -\epsilon \cos kx - k_{3} \cos kx
\end{array} \right |, \qquad
I_{4}=i \left | \begin{array}{c}
-k_{2} \cos kx    \\
  0 \\
  \epsilon \cos kx - k_{3} \cos kx \\
-k_{1} \cos kx - M \sin kx
\end{array} \right |.
\label{6.13}
\end{eqnarray}

Now, let us specify squared solutions in Majorana basis, when starting
with the  real scalar function ($kx = \epsilon t - \vec{k}\vec{x})$):
\begin{eqnarray}
\Phi =  \cos kx , \qquad (i\gamma_{M}^{a} \partial_{a} + M) \cos
kx  = \{\Phi_{j} \}\hspace{30mm}
\nonumber
\\[3mm]
 =
 \left | \begin{array}{cccc}
k_{1}\sin kx  + M \cos kx  &  - \epsilon \sin kx -  k_{3} \sin kx & 0  & - k_{2} \sin kx \\
 \epsilon \sin kx  - k_{3} \sin kx & - k_{1} \sin kx + M \cos kx  & k_{2} \sin kx  & 0\\
0 & k_{2} \sin kx  &  k_{1} \sin  kx   + M\cos kx  &   \epsilon \sin kx - k_{3}  \sin kx \\
 -k_{2} \sin kx  & 0  & - \epsilon \sin kx - k_{3} \sin kx   & -k_{1}\sin kx    +M\cos kx
\end{array}\right | ;
\nonumber
\\
\label{6.14a}
\end{eqnarray}

\noindent note identity
$
\{ \Phi_{j}\}  = \{  R_{j} \} $ -- see (\ref{6.12}).

Similarly, one derives
\begin{eqnarray}
\Phi' = -i \; \sin kx , \qquad (i\gamma_{M}^{a} \partial_{a} + M)
\sin kx  = \{\Phi'_{j} \} \hspace{30mm}
\nonumber
\\[3mm]
=
i \left | \begin{array}{cccc}
k_{1}\cos kx  -M \sin kx  & - \epsilon \cos kx - k_{3} \cos kx & 0  & -k_{2} \cos kx \\
 \epsilon \cos kx  - k_{3} \cos kx & -k_{1} \cos kx -M \sin kx  & k_{2} \cos kx  & 0\\
0 & k_{2} \cos kx  & k_{1} \cos kx   -M\sin kx  &  \epsilon \cos kx -k_{3}  \cos kx \\
- k_{2} \cos kx  & 0  & -\epsilon \cos kx - k_{3} \cos kx   &
-k_{1}\cos kx    -M\sin kx
\end{array}\right | ;
\nonumber
\\
\label{6.15a}
\end{eqnarray}

\noindent again note identity
$
\{ \Phi'_{j} \}  = \{  I_{j} \}$.

It is readily verified, that these two 4-dimensional spaces, $R_{j}$ and $I_{j}$,  cannot
be connected by any linear transformation. To this end, one should
check  the relationships
\begin{eqnarray}
I_{(j)} = S_{(j)k}  R_{(k)}   \quad \Longleftrightarrow\quad
I_{(j)l}
= S_{(j)k}  R_{(k)l} \;.
\label{question}
\end{eqnarray}
Indeed,  from (\ref{question}) we infer that
 the matrix  $S$ depends on coordinates
$$
 I   =   S \;  R \qquad \Longrightarrow \qquad  S=  I \;
R^{-1} = {i \over M^{2}}
$$
$$
\hspace{-5mm} \times \left | \begin{array}{cccc}
Mk_{1}- F \sin 2kx & -(\epsilon+k_{3})^{2}\sin 2kx  &  0  &-k_{2} \sin 2kx    \\
 -(\epsilon - k_{3})^{2}\sin 2kx & -(Mk_{1}+F \sin 2kx) & -k_{2} \sin 2kx   &  0 \\
0& -k_{2} \sin 2kx    &   Mk_{1}-{1 \over 2}(M^{2}+k_{1}^{2}) \sin 2kx &  -(\epsilon-k_{3})^{2}\sin 2kx \\
-k_{2} \sin 2kx  &  0 & -(\epsilon+k_{3})^{2}\sin 2kx   & -
(Mk_{1}+F \sin 2kx)
\end{array} \right |,
$$

\noindent where for brevity we use notation $ {1 \over
2}(M^{2}+k_{1}^{2}) = F $.  It means that these two sets, $\{I_{j}\}$ and
$\{ R_{j}\}$,  determine substantially different spaces: from physical
standpoint they refer respectively to  Majorana
fermions with different properties with respect to the charge
conjugation.

In the charged Dirac case,  at fixed momentum there exist sets of
two states different in helicity, whereas in Majorana real case we
face sets of four solutions only.

\section{Vanishing the current а $J^{z}$  on the boundaries of the domain between two planes
}

Using Dirac plane waves with fixed polarization states  (for
brevity the general factor  $e^{-i \epsilon t} \; e^{i k_{1}x} \;
e^{ik_{2}y}$  is omitted):
\begin{eqnarray}
\Psi_{1} = \Psi _{\alpha, k }  =   e^{ik z}
\left | \begin{array}{c}   1 \\ {k_{1} + i k_{2} \over k  + p}  \\   \alpha   \\
\alpha {k_{1} + i k_{2} \over k  + p}
\end{array} \right |  ,\qquad
\Psi_{2} = \Psi _{(\beta), k}  =  e^{ik z}
\left | \begin{array}{c}   1  \\ {k_{1} + i k_{2} \over k  - p}  \\   \beta   \\
\beta {k_{1} + i k_{2} \over k  - p}
\end{array} \right | \; ,
\label{7.1}
\end{eqnarray}

\noindent and similar ones t with the change  $k \Longrightarrow - k$:
\begin{eqnarray}
\Psi_{3} = \Psi _{\alpha, -k }  =  e^{-ik z}
\left | \begin{array}{c}   1 \\ -{k_{1} + i k_{2} \over k  - p}  \\   \alpha   \\
- \alpha {k_{1} + i k_{2} \over k  - p}
\end{array} \right | ,\qquad
\Psi_{4} = \Psi _{(\beta), -k}  = e^{-ik z}
\left | \begin{array}{c}   1  \\ - {k_{1} + i k_{2} \over k  + p}  \\   \beta   \\
- \beta {k_{1} + i k_{2} \over k  + p}
\end{array} \right | \; ,
\label{7.2}
\end{eqnarray}

\noindent let us make a linear combination
$
\Phi = A_{1} \Psi_{1} + A_{2} \Psi_{2} + A_{3} \Psi_{3} + A_{4}
\Psi_{4} \; .
$ Four components of this wave function  $\Phi$ are
$$
\Phi_{1} = e^{-i\epsilon t} e^{ia x} \left [ e^{ikz}  (A_{1}
+A_{2} ) + e^{-ikz} (A_{3} +A_{4} )  \right ]  ,
$$
$$
\Phi_{2} = e^{-i\epsilon t} e^{ia x}  (k_{1} + i k_{2}) \left [
e^{ikz}  (  {A_{1}  \over k  + p}   + {A_{2}  \over k  - p}  ) -
 e^{-ikz}( {A_{3} \over k  - p}  + {A_{4} \over k+p }) \right ]  ,
$$
$$
\Phi_{3} = e^{-i\epsilon t} e^{ia x} \left [e^{ikz}  (A_{1}\alpha
+A_{2} \beta ) + e^{-ikz} (A_{3}\alpha  +A_{4} \beta ) \right ] ,
$$
\begin{eqnarray}
\Phi_{4} =
 e^{-i\epsilon t} e^{ia x}  (k_{1} + i k_{2}) \left [
e^{ikz}  (  {A_{1} \alpha  \over k  + p}   + {A_{2} \beta  \over k
- p}  ) -
 e^{-ikz}( {A_{3} \alpha \over k  - p}  + {A_{4}\beta  \over k+p }) \right ] .
\label{7.4}
\end{eqnarray}

Note the structure of the current
(in spinor basis) is
$$
J^{z} = \Phi^{+} \gamma^{0} \gamma^{3} \Phi = ( \Phi_{1}^{*}
\Phi_{1} -  \Phi_{3}^ {*} \Phi_{3} ) - ( \Phi_{2}^{*} \Phi_{2} -
\Phi_{4}^{*} \Phi_{4} )\; .
$$

\noindent The current on the boundaries  ($z=-a,+a$) vanishes if the following requirements are fulfilled:
$$
\underline{z = -a}\;, \qquad \Phi_{3} = e^{i\rho } \Phi_{1},
\qquad  \Phi_{4} = e^{i\sigma} \Phi_{2} \qquad \Longrightarrow
$$
\begin{eqnarray}
 \left [e^{-ika}  (A_{1}\alpha  +A_{2} \beta ) + e^{ika} (A_{3}\alpha  +A_{4} \beta ) \right ]=
 e^{i\rho} \left [ e^{-ika}  (A_{1} +A_{2} ) + e^{ika} (A_{3} +A_{4} )  \right ],
\nonumber
\\
\left [ e^{-ika}  (  {A_{1} \alpha  \over k  + p}   + {A_{2} \beta
\over k  - p}  ) -
 e^{ika}( {A_{3} \alpha \over k  - p}  + {A_{4}\beta  \over k+p }) \right ]
 =
e^{i\sigma} \left [ e^{-ika}  (  {A_{1}  \over k  + p}   + {A_{2}
\over k  - p}  ) -
 e^{ika}( {A_{3} \over k  - p}  + {A_{4} \over k+p }) \right ];
\nonumber
\label{7.6a}
\end{eqnarray}

$$
\underline{z = +a}\;, \qquad \Phi_{3} = e^{i\mu } \Phi_{1}, \qquad
\Phi_{4} = e^{i\nu} \Phi_{2} \qquad \Longrightarrow
$$
\begin{eqnarray}
 \left [e^{ika}  (A_{1}\alpha  +A_{2} \beta ) + e^{-ika} (A_{3}\alpha  +A_{4} \beta ) \right ]=
 e^{i\mu } \left [ e^{ika}  (A_{1} +A_{2} ) + e^{-ika} (A_{3} +A_{4} )  \right ],
\nonumber
\\
\left [ e^{ika}  (  {A_{1} \alpha  \over k  + p}   + {A_{2} \beta
\over k  - p}  ) -
 e^{-ika}( {A_{3} \alpha \over k  - p}  + {A_{4}\beta  \over k+p }) \right ]
  =
e^{i\nu} \left [ e^{ika}  (  {A_{1}  \over k  + p}   + {A_{2}
\over k  - p}  ) -
 e^{-ika}( {A_{3} \over k  - p}  + {A_{4} \over k+p }) \right ] .
 \nonumber
 \label{7.6b}
 \end{eqnarray}

Thus, we arrive at the homogeneous linear system with respect to complex
variables
 $A_{1}, A_{2}, A_{3}, A_{4}$.
 It is convenient to introduce notation
 $K=e^{2iak}$, then the above system   reads as
$$
A_{1} (\alpha - e^{i\rho})   +A_{2} (\beta  - e^{i\rho})   + A_{3}
(\alpha - e^{i\rho}) K  +A_{4} (\beta  - e^{i\rho}) K = 0 \; ,
$$
$$
A_{1}K (\alpha -e^{i\mu} )   +A_{2} K  (\beta-e^{i\mu})  +
A_{3}(\alpha -e^{i\mu}) +A_{4} ( \beta  -e^{i\mu}) = 0 \; ,
$$
$$
  A_{1} (\alpha -e^{i\sigma})  (k -p)  + A_{2} (\beta -e^{i\sigma}) (k+p)  -
 A_{3} K (\alpha-e^{i\sigma}) (k+p )- A_{4} K (\beta -e^{i\sigma})( k-p) =0\; ,
$$
$$
  A_{1} K (\alpha -e^{i\nu}) (k -p)  +A_{2} K  (\beta -e^{i\nu})(k+p )  -
 A_{3} (\alpha -e^{i\nu}) (k+p) - A_{4} ( \beta -e^{i\nu}) (k-p ) = 0\;.
$$
\begin{eqnarray}
\label{7.7c}
\end{eqnarray}

Let us write down an explicit form of the main matrix for linear system (\ref{7.7c})
\begin{eqnarray}
\left | \begin{array}{rrrr}
 (\alpha - e^{i\rho})  &  (\beta  - e^{i\rho})   & (\alpha - e^{i\rho}) K  &  (\beta  - e^{i\rho}) K \\
 (\alpha -e^{i\mu} )K  &    (\beta-e^{i\mu}) K & (\alpha -e^{i\mu}) & ( \beta  -e^{i\mu})  \\
(\alpha -e^{i\sigma})  (k +p)  & (\beta -e^{i\sigma}) (k-p)  & -
   (\alpha-e^{i\sigma}) (k-p )K & -  (\beta -e^{i\sigma})( k+p)K \\
    (\alpha -e^{i\nu}) (k +p)K  &   (\beta -e^{i\nu})(k-p )K &  -
  (\alpha -e^{i\nu}) (k-p)  & - ( \beta -e^{i\nu}) (k+p )
  \end{array} \right |.
 \label{7.8}
 \end{eqnarray}

Eq.   $\det S =0$  reduces to a 4-th order polynomial with respect to $K$,
from physical point of view we are interested in roots  which are complex number
$K= e^{2ik a}$ of the unit length -- see the detailed analysis of  its solutions in \cite{3}.

\section{ Vanishing the current on the boundaries,  basis produced by  the squaring method
}

Now let us derive explicit form of vanishing the current $J^{z}$ on the boundaries
of the domain, when using the set of solutions obtained within  the squaring method.
 We start with (the general  factor
 $e^{-i\epsilon t} e^{ik_{1}x} e^{ik_{
2}y}$ is omitted)
\begin{eqnarray}
\Psi_{1} =  \left |
\begin{array}{c}
M\sin kz  \\
0 \\
(\epsilon \sin kz  + i k \cos kz )\\
-(k_{1} + i k_{2}) \sin kz
\end{array} \right |,\quad
\Psi_{2} = \left |
\begin{array}{c}
 0 \\
  M \sin kz  \\
- (k_{1} - i k_{2}) \sin kz \\
(\epsilon \sin kz  -i k \cos kz )
\end{array} \right |,
\nonumber
\\
\Psi_{3} =  \left |
\begin{array}{c}
(\epsilon \sin kz  - i k \cos kz) \\
(k_{1} + i k_{2}) \sin kz  \\
M \sin kz\\
0
\end{array} \right |,\quad
\Psi_{4} =  \left |
\begin{array}{c}
(k_{1} - i k_{2}) \sin kz\\
(\epsilon  \sin kz + i k  \cos kz)\\
0 \\  M \sin kz
\end{array} \right |.
\label{8.1}
\end{eqnarray}

\noindent four components of  the linear combination
$
\Phi = A_{1} \Psi_{1} + A_{2} \Psi_{2} + A_{3} \Psi_{3} + A_{4}
\Psi_{4}$   are
\begin{eqnarray}
\Phi_{1} = A_{1} M\sin kz + A_{2}0  + A_{3} (\epsilon \sin kz  - i
k \cos kz)+ A_{4} (k_{1} - i k_{2}) \sin kz
 \; ,
\nonumber
\\
\Phi_{2} =A_{1} 0+ A_{2} M \sin kz +
 A_{3}(k_{1} + i k_{2}) \sin kz  + A_{4} (\epsilon  \sin kz + i k  \cos kz)
 \;  \; ,
\nonumber
\\
\Phi_{3} = A_{1} (\epsilon \sin kz  + i k \cos kz ) - A_{2}(k_{1}
- i k_{2}) \sin kz +
 A_{3} M \sin kz+ A_{4} 0 \; ,
\nonumber
\\
\Phi_{4} = -A_{1} (k_{1} + i k_{2}) \sin kz + A_{2}(\epsilon \sin
kz  -i k \cos kz )
 + A_{3} 0 + A_{4}M \sin kz \; .
\label{8.2}
\end{eqnarray}
Remembering on  the current structure
$$
J^{z} = \Phi^{+} \gamma^{0} \gamma^{3} \Phi = ( \Phi_{1}^{*}
\Phi_{1} -  \Phi_{3}^ {*} \Phi_{3} ) - ( \Phi_{2}^{*} \Phi_{2} -
\Phi_{4}^{*} \Phi_{4} )\;.
$$

\noindent we impose restrictions:
$$
\underline{z = -a}\;, \qquad \Phi_{3} = e^{i\rho } \Phi_{1},
\qquad  \Phi_{4} = e^{i\sigma} \Phi_{2} \qquad \Longrightarrow
$$
\begin{eqnarray}
A_{1} (\epsilon \sin ak  - i k \cos ak ) - A_{2}(k_{1} - i k_{2})
\sin ak + A_{3} M \sin ak
\nonumber
\\
=
e^{i\rho } [ A_{1} M\sin ak  +
A_{3} (\epsilon \sin ak  + i k \cos ak) + A_{4} (k_{1} - i k_{2})
\sin ak]\; ,
\nonumber
\\
-A_{1} (k_{1} + i k_{2}) \sin ak + A_{2}( \epsilon \sin ak  +i k
\cos ak ) + A_{4}M \sin ak
\nonumber
\\
=
e^{i\sigma} \Phi_{2} [
  A_{2} M \sin ak + A_{3}(k_{1} + i k_{2}) \sin ak  + A_{4} (\epsilon  \sin ak - i k  \cos ak)]\; ;
\nonumber
\end{eqnarray}
$$
\underline{z = +a}\;, \qquad \Phi_{3} = e^{i\mu } \Phi_{1}, \qquad
\Phi_{4} = e^{i\nu} \Phi_{2} \qquad \Longrightarrow
$$
\begin{eqnarray}
A_{1} (\epsilon \sin ak  + i k \cos ka ) - A_{2}(k_{1} - i k_{2})
\sin ak + A_{3} M \sin ak
\nonumber
\\
 =e^{i\mu } [ A_{1} M\sin ak   +
A_{3} (\epsilon \sin ak  - i k \cos ak)+ A_{4} (k_{1} - i k_{2})
\sin ak] \; ,
\nonumber
\\
-A_{1} (k_{1} + i k_{2}) \sin ak + A_{2}(\epsilon \sin ak  -i k
\cos ak )  + A_{4}M \sin ak
\nonumber
\\
 =
 e^{i\nu} \Phi_{2} [
 A_{2} M \sin ak + A_{3}(k_{1} + i k_{2}) \sin ak  + A_{4} (\epsilon  \sin ak + i k  \cos ak)]\; .
\label{8.5b}
\end{eqnarray}

With the use of elementary   identities
\begin{eqnarray}
\cos ak = { e^{iak} + e^{-iak} \over 2}=  { K + K^{-1} \over 2} ,
\qquad
 \sin  ak = -i { e^{iak} - e^{-iak} \over 2} = -i {K-K^{-1} \over 2} \; .
\nonumber
\end{eqnarray}

\noindent
and the  shortening notations
$$
\epsilon +k =  m \;, \;
\epsilon -k = n\; ,\;
k_{1}+ik_{2} = f \; , \;
k_{1}-i k_{2} = g \; ,\;
 mn = fg  \;, \;  g =f^{*} \; ,
$$
$$
 e^{i\rho} = x, \qquad e^{i\mu}  = y, \qquad e^{i\sigma} = v, \qquad e^{i \nu} = w \; ;
$$

\noindent
we arrive at linear equations
\begin{eqnarray}
-A_{1} [   K  ( m- x M ) -  ( n - x M ) ]
 + A_{2} g (K- 1 ) -
A_{3} [   K ( M- x n ) -  ( M- x m ) ]  + A_{4}
 e^{i\rho}g (K-1 ) =0\; ,
\nonumber
\\
-A_{1}  [  K  ( n -y M ) -  ( m - y M ) ]  + A_{2}g (K- 1 ) -
A_{3}  [ K ( M- y m ) -  ( M- y n ) ]  + A_{4}
 y g (K-1 ) =0\; ,
\nonumber
\\
A_{1} f (K-1)  -  A_{2} [   K ( n - v M )
-  (m - v M ) ] +
 A_{3} v f (K -1 )
- A_{4}  [ K ( M-v n ) -  (M-v m ) ] =0,
\nonumber
\\
\hspace{-3mm}
A_{1} f (K-1)  -  A_{2} [  K ( m - e^{i\nu} M )
-  (n - e^{i\nu} M ) ] +
 A_{3} e^{i\nu} f (K -1 )
- A_{4}  [  K ( M-e^{i\nu} m )  -  (M-e^{i\nu} n) ] =0.
\nonumber
\end{eqnarray}

This linear homogeneous system has solutions if its determinant equals to zero:
\begin{eqnarray}
S '=
\label{8.9}
\end{eqnarray}
\small{$$
\hspace{-8mm}\left | \begin{array}{cccc}
-   K  ( m-xM ) +   n - xM   &   g (K- 1 )  & -
    K ( M-xn ) +   M-x  m    &  x g (K-1 ) \\
-    K  ( n - yM ) +  ( m - y M )  & g (K- 1 ) &  -
  K ( M- y m ) +   M- y n  &
 y g (K-1 ) \\
   f (K-1)  & -    K ( n - v M )
+  m - v M    & v f (K -1 ) & -   K ( M-v n ) +  M-v m   \\
  f (K-1)   & -    K ( m - w M )
+  n - w M    & w f (K -1 ) &-    K ( M-w m )  +  M-w n
\end{array} \right |
$$
}

In the paper \cite{DAN-2014} it was shown that any particular choice in 4-dimensional space of states for the Dirac particle
does nod influence substantially the result of solving the equations $\det S=0$ and
$\det S'=0$: the whole  set of the roots is the same.

\section{  2-component Weyl neutrino
}

The possible way to  specify neutrino wave equation is to
consider the Dirac equation in spinor basis
\begin{eqnarray}
\psi (x) =  \left | \begin{array}{c} \xi (x)   \\ \eta (x)
\end{array} \right |   ,   \xi (x) = \left |
\begin{array}{c} \xi ^{1}  \\ \xi ^{2} \end{array} \right | \; ,\quad
 \eta (x) = \left | \begin{array}{c} \eta _{\dot{1}} \\ \eta
_{\dot{2}}
\end{array} \right |   ,
\gamma ^{a} = \left | \begin{array}{cc}
0  &  \bar{\sigma}^{a}   \\
\sigma ^{a} & 0                \end{array} \right | \; , \label{9.1a}
\end{eqnarray}

\noindent where  $\sigma ^{a} = (I,\; +\sigma ^{k}) \; ,\;
\bar{\sigma }^{a} = (I, \;-\sigma ^{k}) $
\begin{eqnarray}
\sigma^{0} = \left | \begin{array}{cc} 1& 0
\\
0 & 1
\end{array} \right |,\quad
 \sigma^{1} = \left | \begin{array}{cc} 0& 1
\\
1 & 0
\end{array} \right |,\quad
\sigma^{2} = \left | \begin{array}{cc} 0& -i
\\
+i & 0
\end{array} \right |,\quad
\sigma^{3} = \left | \begin{array}{cc} 1& 0
\\
0 & -1
\end{array} \right | .
\nonumber
\end{eqnarray}

\noindent Un this basis, we have two equations
\begin{eqnarray}
i  \sigma ^{a}  \partial _{a }
   \xi (x) =   M  \eta (x)    , \quad
i  \bar{\sigma }^{a} \partial _{\alpha }   \eta (x) =  M
\xi (x)  , \label{9.2}
\end{eqnarray}

\noindent setting here $m=0$ we produce equation for neutrino and anti-neutrino:
\begin{eqnarray}
(\mbox{neutrino})\;\; i\; \bar{\sigma }^{a} \partial _{\alpha } \;
\eta (x) = 0\;  ,\quad
(\mbox{anti-neutrino})\;\; i\;
\sigma ^{a}
\partial _{a }
  \; \xi (x) =  0 \;    .
\label{9.3}
\end{eqnarray}

\noindent The neutrino equation differently reads
\begin{eqnarray}
( i\partial_{t} - i \partial_{j} \sigma^{j} ) \eta =0,
\quad
\Sigma \; \eta = - i\partial_{t} \eta,
\label{9.4}
\end{eqnarray}

\noindent
where  $ p_{j}
\sigma^{j}  = \Sigma $ stands for helicity operator.
For the plane wave
\begin{eqnarray}
\eta = e^{-i\epsilon t} e^{i \vec{k}  \vec{x}} \left |
\begin{array}{c}
\eta_{1}\\
\eta_{2}
\end{array}  \right |
\label{9.5a}
\end{eqnarray}

\noindent from (\ref{9.4}) it follows that helicity of the neutrino with fixed momentum is
automatically  negative:
$$
\Sigma \; \eta= - \epsilon \; \eta\;, \qquad \epsilon
=+\sqrt{k_{1}^{2} + k_{2}^{2} + k_{3}^{2}} \; .
$$

\noindent  For those states one finds restriction on the components
 $\eta_{1},
\eta_{2}$  in  (\ref{9.5a}):
\begin{eqnarray}
\left | \begin{array}{cc}
  \epsilon +k_{3}  & k_{1}  - ik_{2} \\
 k_{1}  +i k_{2}     & \epsilon -k_{3}
      \end{array} \right |\left | \begin{array}{c}
\eta_{1}\\
\eta_{2}
\end{array}  \right |=0 \; \Longrightarrow \;
\epsilon^{2}= k_{1}^{2} + k_{2}^{2} + k_{3}^{2}\;, \;
\eta_{2} = - {k_{1} + i k_{2} \over \epsilon - k_{3}}\; \eta_{1}\; .
\label{9.5c}
\end{eqnarray}

Let the neutrino plane waves  be  written in the form
(for simplicity,  $\eta_{1}=1$; and  $k _{3}= k$; we will need also
solutions with opposite $k$)
\begin{eqnarray}
\eta = e^{-i\epsilon t} e^{i xk_{1} } e^{i yk_{2} } e^{i z k}
\left | \begin{array}{c}
1\\[3mm]
- {k_{1} + i k_{2} \over \epsilon - k}
\end{array}  \right | ,\quad
\eta' = e^{-i\epsilon t} e^{i xk_{1} } e^{i yk_{2} } e^{-i z k}
\left | \begin{array}{c}
1\\[3mm]
- {k_{1} + i k_{2} \over \epsilon + k}
\end{array}  \right |.
\label{9.6}
\end{eqnarray}

Summing equations for  $\eta$  and  $\eta^{+}$:
\begin{eqnarray}
\eta^{+}( \partial_{t} -  \stackrel{\rightarrow}{\partial}_{j}
\sigma^{j} ) \eta =0\;, \; \eta^{+} (
\stackrel{\leftarrow}{\partial}_{t} -  \partial_{j} \sigma^{j} )
\eta=0
\nonumber
\end{eqnarray}
one derives expression for the neutrino current
\begin{eqnarray}
\partial_{a} J^{a} =0 \; , \;
 J^{a} = \left (\eta^{+} \eta , \;   - \eta^{+} \sigma^{j} \eta  \right ) .
\label{7a}
\end{eqnarray}
Its components are
\begin{eqnarray}
J^{t} = \eta_{1}^{*}  \eta _{1} +  \eta_{2}^{*}  \eta _{2} \;,
\quad J^{z} = - \eta_{1}^{*}  \eta _{1} +  \eta_{2}^{*}  \eta
_{2} \;,
\nonumber
\\
J^{1} =-  \eta_{1}^{*}  \eta _{2} -  \eta_{2}^{*}  \eta _{1}
\;, \qquad J^{2} =i    \eta_{1}^{*}  \eta _{2} -  i
\eta_{2}^{*}  \eta _{1}  \;.
\nonumber
\label{9.7b}
\end{eqnarray}

In particular, the component $j^{z} $  is given by
\begin{eqnarray}
j^{z} =  -\left ( 1 - {k_{1}^{2} +k_{2}^{2} \over (\epsilon - k)^{2} } \right )={k \over \epsilon - k}\;.
\label{9.8}
\end{eqnarray}

Now, let us examine  the vanishing of the current $J^{z}$ on the boundaries
of the domain between two planes.
The structure of the current indicates the way to reach this:
\begin{eqnarray}
J^{z} = - \eta_{1}^{*} \; \eta _{1} +  \eta_{2}^{*} \; \eta
_{2} \;\;\Longrightarrow \;\; \eta_{2} = e^{i\gamma} \eta_{1} \;
. \label{9.9}
\end{eqnarray}

\noindent To satisfy this requirement, let us introduce a  special linear combinations of the
plane waves with opposite momentums (general factor is omitted)
\begin{eqnarray}
H = A \eta + B \eta '
 =   \left |
\begin{array}{c}
A e^{i z k} + B e^{-i z k} \\[2mm]
- {k_{1} + i k_{2} \over \epsilon - k}A e^{i z k}
- {k_{1} + i k_{2} \over \epsilon + k}B e^{-i z k}
\end{array}  \right |
\nonumber
 \label{9.10}
\end{eqnarray}

\noindent
for which the above current  restrictions read
$$
z=-a: \qquad  \qquad H_{2} (-a) = e^{i\rho} \; H_{1}(-a)\; ,
$$
$$
- {k_{1} + i k_{2} \over \epsilon - k}A e^{-ia k}
- {k_{1} + i k_{2} \over \epsilon + k}B e^{i a k}
=e^{i\rho} [ A e^{-i a k} + B e^{i a k} ]\; ;
$$
$$
z=+a: \qquad  \qquad H_{2} (a) = e^{i\sigma}\;  H_{1}(a) \;,
$$
$$
- {k_{1} + i k_{2} \over \epsilon - k}A e^{ia k}
- {k_{1} + i k_{2} \over \epsilon + k}B e^{-i a k}
=e^{i\sigma}  [ A e^{i a k} + B e^{-i a k} ].
$$

Let us introduce parameter  $e^{2iak} = K $,  then
we get the linear homogeneous system
 with respect to $A$ an $B$:
 \begin{eqnarray}
 A   (e^{i\rho}   + {k_{1} + i k_{2} \over \epsilon - k}  )
 + B K  (    e^{i\rho}  + {k_{1} + i k_{2} \over \epsilon + k}   ) = 0 \;,
\nonumber
\\
A  K  (
e^{i\sigma}      + {k_{1} + i k_{2} \over \epsilon - k}   )
 + B   ( e^{i\sigma}  + {k_{1} + i k_{2} \over \epsilon + k} )=0\; .
\nonumber
\label{9.11b}
\end{eqnarray}

\noindent With the shortening notations
$$
{k_{1} + i k_{2} \over \epsilon - k}= f \; ,\quad
{k_{1} + i k_{2} \over \epsilon + k} =g \; ,
\quad  fg^{*}=1 \; , \quad  f^{*}  g = 1 \; ;
$$
 the system reads
\begin{eqnarray}
 A  \left (e^{i\rho}   + f \right )
 + B K \left (    e^{i\rho}  + g  \right ) = 0\; ,
\nonumber
\\
A  K \left (
e^{i\sigma}      + f \right )
 + B  \left ( e^{i\sigma}  +g  \right )=0 \;.
\label{9.11c}
\end{eqnarray}

\noindent
Nontrivial solution exists if $K^{2}$ satisfies the equation below
\begin{eqnarray}
K^{2} =  {  (e^{i\rho}+f)(e^{i\sigma}+g) \over (e^{i\sigma}+f) (e^{i\rho}+g) }  \; .
\label{9.12a}
\end{eqnarray}

\noindent
Let us simplify the notation:
$
e^{i\rho}=x,\;\; e^{i\sigma}= y,$
then
\begin{eqnarray}
K^{2} =  {  (x +f)( y +g) \over (x +g) (y +f)}   \; ;
\label{9.13a}
\end{eqnarray}

\noindent
It is ready proved that the right hand part of (\ref{9.13a}) indeed is a complex number of the  unit length.
To this end,
 it suffices to multiply it by  the complex conjugate -- so we obtain the needed
identity
\begin{eqnarray}
  {  (x +f)( y +g) \over (x +g) (y +f)}   \;  {  (1/x +f^{*})( 1/y +g^{*}) \over (1/x +g^{*}) (1/y +f^{*})}\qquad
  =
{  (x +f)( y +g) \over (x +g) (y +f)}\; {(g+x) \over (f+x) }{ x f  \over x g}\; {(y+f)  \over (y+ g)} {  yg \over y f}
\equiv 1 \; .
\nonumber
\end{eqnarray}

\noindent
then  eq.  (\ref{9.13a}) reads as the quantization rule for $k$ in the form
\begin{eqnarray}
e^{4iak}  =   {  (e^{i\rho} +f)( e^{i\sigma}  +g) \over (e^{i\rho}  +g) (e^{i \sigma}  +f)}    \; .
\label{9.13b}
\end{eqnarray}
This equation permits different solutions. Mostly of them can be
found only numerically. For the case  $e^{i\rho} = e^{i \sigma}$,
we have an analytically solvable equation, $e^{4iak} =1$.

\section{ On the representation of   $J^{z}=0$ in arbitrary basis of the  Dirac matrices}

In spinor basis, vanishing of the current is achieved
if
\begin{eqnarray}
\Psi_{3} = e^{i\rho} \Psi_{1}\; , \qquad \Psi_{4} = e^{i \sigma}\Psi_{2}\; ;
\label{10.1a}
\end{eqnarray}

\noindent however these relations are not covariant: they are valid
only in this particular basis.
This deficiency can be resolved.
Let us present (\ref{10.1a}) in the matrix form $\Psi = G\; \Psi $:
\begin{eqnarray}
\left | \begin{array}{c}
\Psi_{1} \\ \Psi_{2} \\ \Psi_{3} \\ \Psi_{4}
\end{array} \right |=
\left | \begin{array}{cccc}
0 & 0 & e^{-i\rho} & 0 \\
0 & 0 & 0 & e^{-i \sigma} \\
e^{i\rho} & 0 & 0 & 0 \\
0 &e^{i\sigma} & 0 & 0
\end{array} \right |
 \left | \begin{array}{c}
\Psi_{1} \\ \Psi_{2} \\ \Psi_{3} \\ \Psi_{4}
\end{array} \right | .
\label{10.1b}
\end{eqnarray}

\noindent
The matrix  $G$ can be decomposed in terms of Dirac matrices
in spinor representation
-- such form will be automatically covariant (valid in all bases)
\begin{eqnarray}
\hspace{-2mm}
\gamma^{a} =
\left | \begin{array}{cc}
0 & \bar{\sigma}^{a} \\
\sigma^{a} & 0
\end{array} \right |  , \gamma^{5} =
\left | \begin{array}{cc}
-I &  0  \\
0  & I
\end{array} \right |,  \gamma^{5}\gamma^{a}=
\left | \begin{array}{cc}
0 & -\bar{\sigma}^{a} \\
\sigma^{a} & 0
\end{array} \right | .
\nonumber
\end{eqnarray}

\noindent Decomposition for $G$  (\ref{10.1b}) should be searched in the form
\begin{eqnarray}
G =(n_{0} \gamma^{0}  +n_{3} \gamma^{3} )  +  \gamma^{5} (m_{0}\gamma^{0}  +m_{3} \gamma^{3} )\; ;
\label{10.2a}
\end{eqnarray}

\noindent which is equivalent to
\begin{eqnarray}
G =
\left | \begin{array}{cc}
0 & n_{0}-m_{0}- (n_{3}-m_{3})\sigma^{3} \\
n_{0}+m_{0} + (n_{3}+m_{3})\sigma^{3} & 0
\end{array} \right |
\nonumber
\end{eqnarray}

\noindent
and  leads to the system
\begin{eqnarray}
(n_{0} -m_{0})- (n_{3} -m_{3}) = e^{-i\rho } \; ,
\nonumber
\\
(n_{0} -m_{0})+ (n_{3} -m_{3}) = e^{-i\sigma } \;  ,
\nonumber
\\
(n_{0} +m_{0}) + (n_{3} +m_{3}) = e^{+i\rho } \; ,
\nonumber
\\
(n_{0} +m_{0}) - (n_{3} +m_{3}) = e^{+i\sigma } \; .
\label{10.3a}
\end{eqnarray}

\noindent Its solution is
\begin{eqnarray}
n_{0} ={1 \over 2} ( \cos \rho + \cos \sigma) \;,
\;n_{3} = {1 \over 2}(i \sin \rho - i \sin \sigma )\; ,
\nonumber
\\
m_{0} ={1 \over 2} (  i \sin \rho + i  \sin \sigma) \;,
\; m_{3} = {1 \over 2}(  \cos \rho - \cos  \sigma )\; .
\nonumber
\label{10.3b}
\end{eqnarray}

Correspondingly,  the matrix $G$    is presented as follows
\begin{eqnarray}
G ={1 \over 2} ( \cos \rho + \cos \sigma) \gamma^{0}
+ {1 \over 2}(i \sin \rho - i \sin \sigma ) \gamma^{3}
\nonumber
\\
+ \gamma^{5} [  {1 \over 2} (  i \sin \rho + i  \sin \sigma) \gamma^{0}
+
{1 \over 2}(  \cos \rho - \cos  \sigma )  \gamma^{3}
 ] \; ;
\nonumber
\label{10.4}
\end{eqnarray}

This can be translated to other  form
\begin{eqnarray}
G =  e^{+i \rho}  \; {1 +\gamma^{5} \over 2 } \; { \gamma^{0} + \gamma^{3} \over 2 }
+
 e^{-i \rho} \; { 1 - \gamma^{5}\over 2 } \; { \gamma^{0} - \gamma^{3}  \over 2 }
 \nonumber
 \\
+  e^{+i \sigma } \; { 1 +\gamma^{5} \over 2 } \; { \gamma^{0} - \gamma^{3} \over 2 }
+
  e^{-i \sigma} \; { 1 - \gamma^{5}\over 2 } \; { \gamma^{0} + \gamma^{3} \over 2 } \; .
\label{10.6}
\end{eqnarray}

Condition for vanishing the current in covariant form is given by the relation
\begin{eqnarray}
J^{z}=0 \qquad \Longleftrightarrow \qquad G  (\rho , \sigma ) \; \Psi = \Psi \; .
\label{result}
\end{eqnarray}

\noindent
Formula (\ref{10.6}) becomes simpler at $e^{i\sigma} = e^{i\rho}$:
\begin{eqnarray}
G =  e^{+i \rho}  \; {1 + \gamma^{5} \over 2 } \;  \gamma^{0}   +
 e^{-i \rho} \; { 1 - \gamma^{5}\over 2 } \;  \gamma^{0} \; ;
\label{10.8}
\end{eqnarray}

\noindent
Similarly,  (\ref{10.6}) is simplified if
 $e^{i\sigma} = e^{-i\rho}$:
\begin{eqnarray}
G =  e^{+i \rho}  \;  { \gamma^{0} + \gamma^{3} \over 2 }  +
 e^{-i \rho} \;  { \gamma^{0} - \gamma^{3}  \over 2 } \; .
 \label{10.9}
 \end{eqnarray}

\section*{10. Conclusion}

The accent in the paper was given to
consideration of solutions of the Dirac equation which
 have vanishing  the third projection of the
conserved current $J^{z}$ on the boundaries of the domain between two  parallel planes. Such solutions
are reachable when considering 4-dimensional   basis of four
solutions -- plane waves
 with opposite signs  of the third projection of
momentum
 $+k_{3}$ and   $-k_{3}$.
It is  shown that the known  method  to solve the Dirac equation
trough squaring method, if based on the  scalar
function
 $
 \Phi = e^{-i\epsilon t} e^{ik_{1}x} e^{ik_{2}y} \sin (kz
+ \alpha)$,
 leads  to 4-dimensional basis of Dirac solutions. It is shown that
so constructed basis  is equivalent to the 4-space of Dirac plane
wave states;
 corresponding linear transformation is found.
Application  of the squaring method in Majorana representation is
investigated as well.

 General conditions for vanishing  third projection of
the  current $J^{z}$
 at the boundaries of the domain between two parallel planes are formulated:
 on the 4-dimensional base of plane spinor waves and  on the base of the   squared basis.
In both cases, these conditions  reduce to linear homogeneous
algebraic systems, equation $\det S =0$  for which turns  to be  a
4-th
 order polynomial, the roots
of which are
 $e^{2ik a}$,  where $a$ is a half-distance between the planes.
Solutions of this 4-th order polynomial will be considered in a separate paper.

The case of Weyl neutrino field is investigated as well; conditions for  vanishing the neutrino
current $J^{z}$ on the two boundaries reduce to 2-nd order polynomial: its solutions are found; all of them
are complex numbers with the unit length.

\vspace{5mm}

Authors are grateful to Yu.A. Sitenko for discussion  and advices.

\vspace{5mm}

This  work was   supported   by the Fund for Basic Researches of Belarus,
 F 13K-079, within the cooperation framework between Belarus  and Ukraine,
 and by the   Ministry of Education of Republic of Belarus
for  work on probation in Bogolyubov Institute for Theoretical Physics,
National Academy of Sciences of Ukraine.

\end{document}